\pdfoutput=1
\documentclass[12pt,a4paper]{article}

\usepackage{ifthen} 
\newboolean{pdflatex}
\setboolean{pdflatex}{true} 

\newboolean{articletitles}
\setboolean{articletitles}{true} 

\newboolean{uprightparticles}
\setboolean{uprightparticles}{false} 

\def\paperauthors{LHCb collaboration} 
\def\paperasciititle{Measurement of antiproton
production from antihyperon decays in pHe collisions at sqrt(sNN)=110 GeV} 
\def\papertitle{Measurement of antiproton production from antihyperon decays
in \pHe collisions at $\sqsnn=110$\gev}
\def\paperkeywords{{High Energy Physics}, {LHCb}} 
\def\papercopyright{\the\year\ CERN for the benefit of the LHCb collaboration} 
\def\paperlicence{CC BY 4.0 licence}
\def\paperlicenceurl{https://creativecommons.org/licenses/by/4.0/}

\usepackage[top=1in, bottom=1.25in, left=1in, right=1in]{geometry}

\usepackage{microtype}
\usepackage{lineno}  
\usepackage{xspace} 
\usepackage{caption} 

\usepackage{graphicx}  
\usepackage{color}
\usepackage{colortbl}
\graphicspath{{./figs/}} 

\usepackage{amsmath} 
\usepackage{amssymb}
\usepackage{amsfonts}
\usepackage{upgreek} 

\newcommand*\patchAmsMathEnvironmentForLineno[1]{%
\expandafter\let\csname old#1\expandafter\endcsname\csname #1\endcsname
\expandafter\let\csname oldend#1\expandafter\endcsname\csname
end#1\endcsname
 \renewenvironment{#1}%
   {\linenomath\csname old#1\endcsname}%
   {\csname oldend#1\endcsname\endlinenomath}%
}
\newcommand*\patchBothAmsMathEnvironmentsForLineno[1]{%
  \patchAmsMathEnvironmentForLineno{#1}%
  \patchAmsMathEnvironmentForLineno{#1*}%
}
\AtBeginDocument{%
\patchBothAmsMathEnvironmentsForLineno{equation}%
\patchBothAmsMathEnvironmentsForLineno{align}%
\patchBothAmsMathEnvironmentsForLineno{flalign}%
\patchBothAmsMathEnvironmentsForLineno{alignat}%
\patchBothAmsMathEnvironmentsForLineno{gather}%
\patchBothAmsMathEnvironmentsForLineno{multline}%
\patchBothAmsMathEnvironmentsForLineno{eqnarray}%
}

\usepackage{hyperxmp}

\usepackage[pdftex,
            pdfauthor={\paperauthors},
            pdftitle={\paperasciititle},
            pdfkeywords={\paperkeywords},
            pdfcopyright={Copyright (C) \papercopyright},
            pdflicenseurl={\paperlicenceurl}]{hyperref}

\usepackage[colorinlistoftodos,textsize=scriptsize]{todonotes}

\usepackage[bottom,flushmargin,hang,multiple]{footmisc}

\usepackage[all]{hypcap} 

\usepackage{xspace} 
\usepackage{upgreek}

\def\lhcb   {\mbox{LHCb}\xspace}

\def\lhc    {\mbox{LHC}\xspace}

\def\velo   {VELO\xspace}
\def\rich   {RICH\xspace}

\def\MagUp {\mbox{\em Mag\kern -0.05em Up}\xspace}

\ifthenelse{\boolean{uprightparticles}}%
{

 \def\Ppi         {\ensuremath{\uppi}\xspace}

 \def\PDelta      {\ensuremath{\Delta}\xspace}                 
 \def\PXi         {\ensuremath{\Xi}\xspace}                 
 \def\PLambda     {\ensuremath{\Lambda}\xspace}                 
 \def\PSigma      {\ensuremath{\Sigma}\xspace}                 
 \def\POmega      {\ensuremath{\Omega}\xspace}                 
 \def\PUpsilon    {\ensuremath{\Upsilon}\xspace}
 \let\oldPi\Pi
 \def\PPi         {\ensuremath{\oldPi}\xspace}

 \def\PB      {\ensuremath{\mathrm{B}}\xspace}                 
                  
 \def\PD      {\ensuremath{\mathrm{D}}\xspace}

 \def\PK      {\ensuremath{\mathrm{K}}\xspace}

 \def\Pb      {\ensuremath{\mathrm{b}}\xspace}                 
 \def\Pc      {\ensuremath{\mathrm{c}}\xspace}

 \def\Pi      {\ensuremath{\mathrm{i}}\xspace}

 \def\Pp      {\ensuremath{\mathrm{p}}\xspace}

 \def\Ps      {\ensuremath{\mathrm{s}}\xspace}

 \def\thebaroffset{0.0em}
}
{

 \def\Ppi         {\ensuremath{\pi}\xspace}

 \mathchardef\PDelta="7101
 \mathchardef\PXi="7104
 \mathchardef\PLambda="7103
 \mathchardef\PSigma="7106
 \mathchardef\POmega="710A
 \mathchardef\PUpsilon="7107
 \mathchardef\PPi="7105
                  
 \def\PB      {\ensuremath{B}\xspace}                 
                  
 \def\PD      {\ensuremath{D}\xspace}

 \def\PK      {\ensuremath{K}\xspace}

 \def\Pb      {\ensuremath{b}\xspace}                 
 \def\Pc      {\ensuremath{c}\xspace}

 \def\Pi      {\ensuremath{i}\xspace}

 \def\Pp      {\ensuremath{p}\xspace}

 \def\Ps      {\ensuremath{s}\xspace}

 \def\thebaroffset{0.18em}
}
\newcommand{\offsetoverline}[2][\thebaroffset]{\kern #1\overline{\kern -#1 #2}}%

\makeatletter
\ifcase \@ptsize \relax
  \newcommand{\miniscule}{\@setfontsize\miniscule{4}{5}}
\or
  \newcommand{\miniscule}{\@setfontsize\miniscule{5}{6}}
\or
  \newcommand{\miniscule}{\@setfontsize\miniscule{5}{6}}
\fi
\makeatother

\DeclareRobustCommand{\optbar}[1]{\shortstack{{\miniscule (\rule[.5ex]{1.25em}{.18mm})}
  \\ [-.7ex] $#1$}}

\def\squark    {{\ensuremath{\Ps}}\xspace}

\def\cquark    {{\ensuremath{\Pc}}\xspace}
\def\cquarkbar {{\ensuremath{\overline \cquark}}\xspace}
\def\ccbar     {{\ensuremath{\cquark\cquarkbar}}\xspace}
\def\bquark    {{\ensuremath{\Pb}}\xspace}

\def\pion   {{\ensuremath{\Ppi}}\xspace}
\def\piz    {{\ensuremath{\pion^0}}\xspace}
\def\pip    {{\ensuremath{\pion^+}}\xspace}
\def\pim    {{\ensuremath{\pion^-}}\xspace}

\def\kaon    {{\ensuremath{\PK}}\xspace}

\def\KorKbar {\kern \thebaroffset\optbar{\kern -\thebaroffset \PK}{}\xspace}

\def\Kp      {{\ensuremath{\kaon^+}}\xspace}

\def\KS      {{\ensuremath{\kaon^0_{\mathrm{S}}}}\xspace}

\def\D       {{\ensuremath{\PD}}\xspace}

\def\DorDbar {\kern \thebaroffset\optbar{\kern -\thebaroffset \PD}\xspace}

\def\Dp      {{\ensuremath{\D^+}}\xspace}
\def\Dm      {{\ensuremath{\D^-}}\xspace}

\def\DpDm    {\ensuremath{\Dp {\kern -0.16em \Dm}}\xspace}

\def\B       {{\ensuremath{\PB}}\xspace}

\def\BorBbar {\kern \thebaroffset\optbar{\kern -\thebaroffset \PB}\xspace}

\def\Bd      {{\ensuremath{\B^0}}\xspace}

\def\BdorBdbar {\kern \thebaroffset\optbar{\kern -\thebaroffset \Bd}\xspace}

\def\Bs      {{\ensuremath{\B^0_\squark}}\xspace}

\def\BsorBsbar {\kern \thebaroffset\optbar{\kern -\thebaroffset \Bs}\xspace}

\def\Y#1S{\ensuremath{\PUpsilon{(#1S)}}\xspace}

\def\proton      {{\ensuremath{\Pp}}\xspace}
\def\antiproton  {{\ensuremath{\overline \proton}}\xspace}

\def\Lz          {{\ensuremath{\PLambda}}\xspace}
\def\Lbar        {{\ensuremath{\offsetoverline{\PLambda}}}\xspace}
\def\LorLbar     {\kern \thebaroffset\optbar{\kern -\thebaroffset \PLambda}\xspace}

\def\Sigmares    {{\ensuremath{\PSigma}}\xspace}

\def\Sigmaresbar {{\ensuremath{\offsetoverline{\Sigmares}}}\xspace}

\def\Sigmabarm   {{\ensuremath{\Sigmaresbar{}^-}}\xspace}
\def\Xires       {{\ensuremath{\PXi}}\xspace}

\def\Xiresbar       {{\ensuremath{\offsetoverline{\Xires}}}\xspace}
\def\Xibarz      {{\ensuremath{\Xiresbar^0}}\xspace}
\def\Xibarp      {{\ensuremath{\Xiresbar^+}}\xspace}

\def\Omegaresbar {{\ensuremath{\offsetoverline{\POmega}}}\xspace}

\def\Omegabarp   {{\ensuremath{\Omegaresbar^+}}\xspace}

\newcommand{\decay}[2]{\ensuremath{#1\!\to #2}\xspace} 

\def\to                 {\ensuremath{\rightarrow}\xspace}

\def\AT#1     {\ensuremath{A_{\mathrm{T}}^{#1}}\xspace}           

\def\C#1      {\ensuremath{\mathcal{C}_{#1}}\xspace}                       
\def\Cp#1     {\ensuremath{\mathcal{C}_{#1}^{'}}\xspace}                    
\def\Ceff#1   {\ensuremath{\mathcal{C}_{#1}^{\mathrm{(eff)}}}\xspace}        
\def\Cpeff#1  {\ensuremath{\mathcal{C}_{#1}^{'\mathrm{(eff)}}}\xspace}       
\def\Ope#1    {\ensuremath{\mathcal{O}_{#1}}\xspace}                       
\def\Opep#1   {\ensuremath{\mathcal{O}_{#1}^{'}}\xspace}                    

\newcommand{\nospaceunit}[1]{\ensuremath{\text{#1}}}       
\newcommand{\aunit}[1]{\ensuremath{\text{\,#1}}}       

\newcommand{\tev}{\aunit{Te\kern -0.1em V}\xspace}
\newcommand{\gev}{\aunit{Ge\kern -0.1em V}\xspace}
\newcommand{\mev}{\aunit{Me\kern -0.1em V}\xspace}
\newcommand{\kev}{\aunit{ke\kern -0.1em V}\xspace}
\newcommand{\ev}{\aunit{e\kern -0.1em V}\xspace}
 
\newcommand{\mevc}{\ensuremath{\aunit{Me\kern -0.1em V\!/}c}\xspace}
\newcommand{\gevc}{\ensuremath{\aunit{Ge\kern -0.1em V\!/}c}\xspace}
\newcommand{\mevcc}{\ensuremath{\aunit{Me\kern -0.1em V\!/}c^2}\xspace}
\newcommand{\gevcc}{\ensuremath{\aunit{Ge\kern -0.1em V\!/}c^2}\xspace}

\def\mm   {\aunit{mm}\xspace}

\def\mum  {\ensuremath{\,\upmu\nospaceunit{m}}\xspace}

\def\nb {\aunit{nb}\xspace}
\def\invnb {\ensuremath{\nb^{-1}}\xspace}

\def\mus  {\ensuremath{\,\upmu\nospaceunit{s}}\xspace}

\newcommand{\chisq}{\ensuremath{\chi^2}\xspace}

\newcommand{\chisqip}{\ensuremath{\chi^2_{\text{IP}}}\xspace}

\def\gsim{{~\raise.15em\hbox{$>$}\kern-.85em
          \lower.35em\hbox{$\sim$}~}\xspace}
\def\lsim{{~\raise.15em\hbox{$<$}\kern-.85em
          \lower.35em\hbox{$\sim$}~}\xspace}

\def\sPlot{\mbox{\em sPlot}\xspace}

\def\sqsnn {\ensuremath{\protect\sqrt{s_{\scriptscriptstyle\text{NN}}}}\xspace}
\def\pt         {\ensuremath{p_{\mathrm{T}}}\xspace}

\def\ptot       {\ensuremath{p}\xspace}

\def\dllkpi     {\ensuremath{\mathrm{DLL}_{\kaon\pion}}\xspace}
\def\dllppi     {\ensuremath{\mathrm{DLL}_{\proton\pion}}\xspace}

\def\geant      {\mbox{\textsc{Geant4}}\xspace}

\def\tell1  {TELL1\xspace}
\def\ukl1   {UKL1\xspace}

\newcommand{\lhcborcid}[1]{\href{https://orcid.org/#1}{\hspace*{0.1em}\raisebox{-0.45ex}{\includegraphics[width=1em]{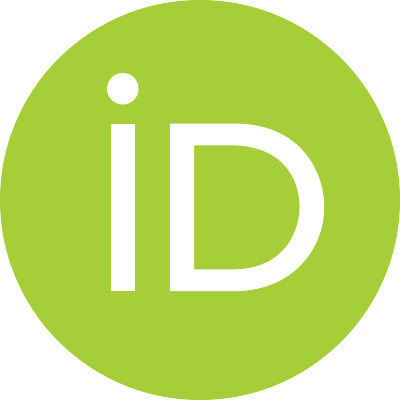}}}}

\usepackage{cite} 
\usepackage{mciteplus}

\newcommand{\ped}[1]{\ensuremath{_{\text{#1}}}\xspace}
\newcommand{\ap}[1]{\ensuremath{^{\text{#1}}}\xspace}

\def\eposlhc{\mbox{\textsc{Epos-lhc}}\xspace}
\def\eposold{\mbox{\textsc{Epos~1.99}}\xspace}
\def\qgsjet{\mbox{\textsc{Qgsjet-ii04}}\xspace}
\def\pythiaold{\mbox{\textsc{Pythia~6}}\xspace}
\def\hijing{\mbox{\textsc{Hijing~1.38}}\xspace}
\def\crmc{\mbox{\textsc{Crmc}}\xspace}

\def\sqsnn{\ensuremath{\protect\sqrt{s_{\scriptscriptstyle\rm NN}}}\xspace}
\def\pHe{\ensuremath{\proton {\rm He}}\xspace}
\def\pNe{\ensuremath{\proton {\rm Ne}}\xspace}
\def\pp{\proton\proton}
\def\pbar{\antiproton}
\def\antihyp{\ensuremath{\offsetoverline{H}}\xspace}
\def\wm{\ensuremath{M_{\proton \rightarrow \pi}}\xspace}

\def\dllpk{\ensuremath{\mathrm{DLL}_{\proton\kaon}}\xspace}

\def\Lbardecay{\mbox{\ensuremath{\Lbar\to\pbar\pip}}\xspace}
\def\Rinc{\ensuremath{R_{\antihyp}}\xspace}
\def\Rexc{\ensuremath{R_{\Lbar}}\xspace}
\def\KSdecay{\ensuremath{\decay{\KS}{\pim \pip}}\xspace}
\def\Lbarstardecay{\ensuremath{\decay{\Lbar(1520)}{\pbar \Kp}}\xspace}
\def\Xibarpdecay{\ensuremath{\decay{\Xibarp}{\Lbar\pip}}\xspace}
\def\Xibarzdecay{\ensuremath{\decay{\Xibarz}{\Lbar\piz}}\xspace}
\def\Sbar{\Sigmabarm}
\def\Sbardecay{\ensuremath{\decay{\Sigmabarm}{\pbar\piz}}\xspace}
\def\fip{\ensuremath{\mathcal{F}\ped{IP}}\xspace}
\def\fipchi{\ensuremath{\mathcal{F}\ped{\chisqip}}\xspace}
\def\chisqdoca{\ensuremath{\chisq_{\text{DOCA}}}\xspace}

\def\Tm    {\aunit{Tm}\xspace}
\def\mbar  {\aunit{mbar}\xspace}

\usepackage{multirow}
\usepackage{booktabs}

\begin{document}

\renewcommand{\thefootnote}{\fnsymbol{footnote}}
\setcounter{footnote}{1}


\begin{titlepage}
\pagenumbering{roman}

\vspace*{-1.5cm}
\centerline{\large EUROPEAN ORGANIZATION FOR NUCLEAR RESEARCH (CERN)}
\vspace*{1.5cm}
\noindent
\begin{tabular*}{\linewidth}{lc@{\extracolsep{\fill}}r@{\extracolsep{0pt}}}
\ifthenelse{\boolean{pdflatex}}
{\vspace*{-1.5cm}\mbox{\!\!\!\includegraphics[width=.14\textwidth]{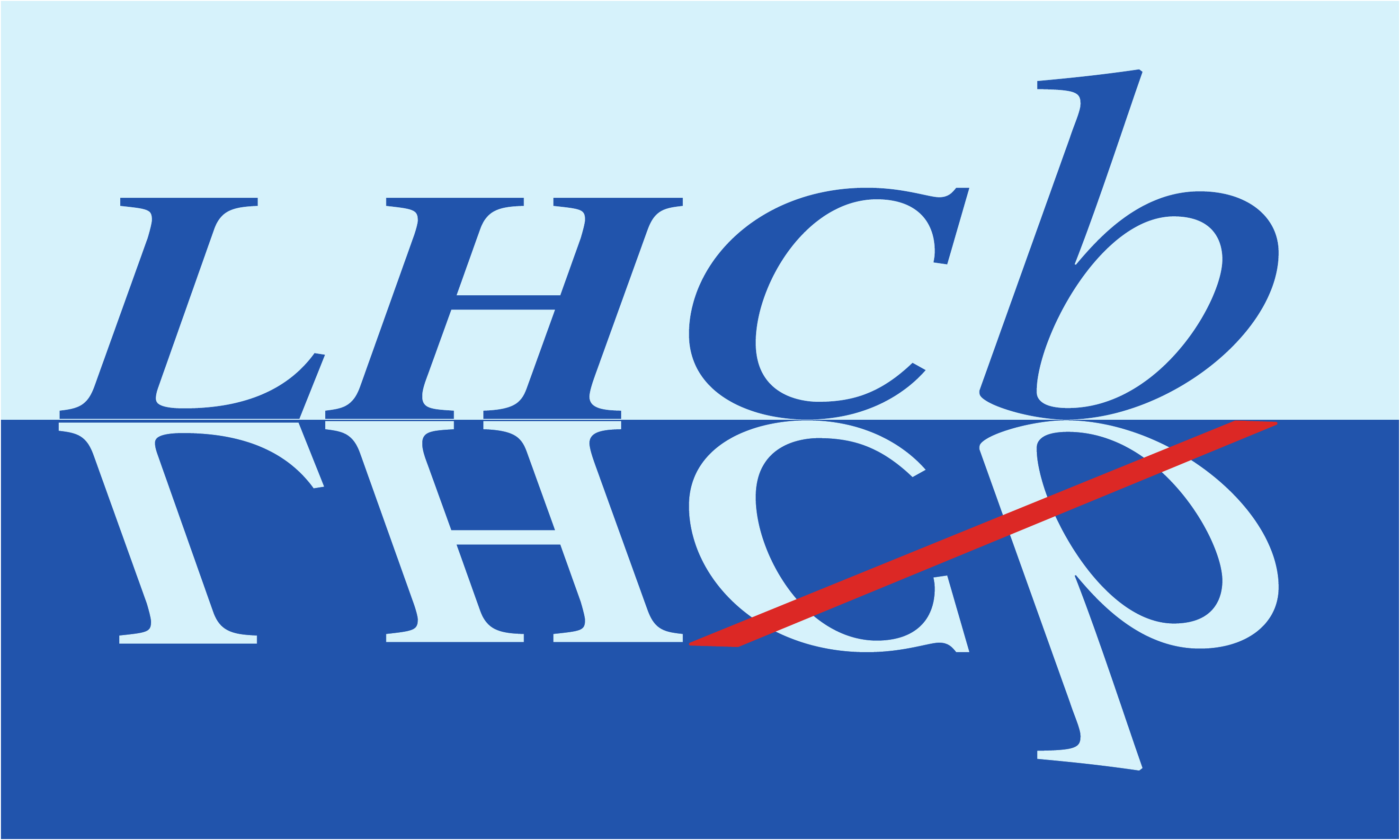}} & &}%
{\vspace*{-1.2cm}\mbox{\!\!\!\includegraphics[width=.12\textwidth]{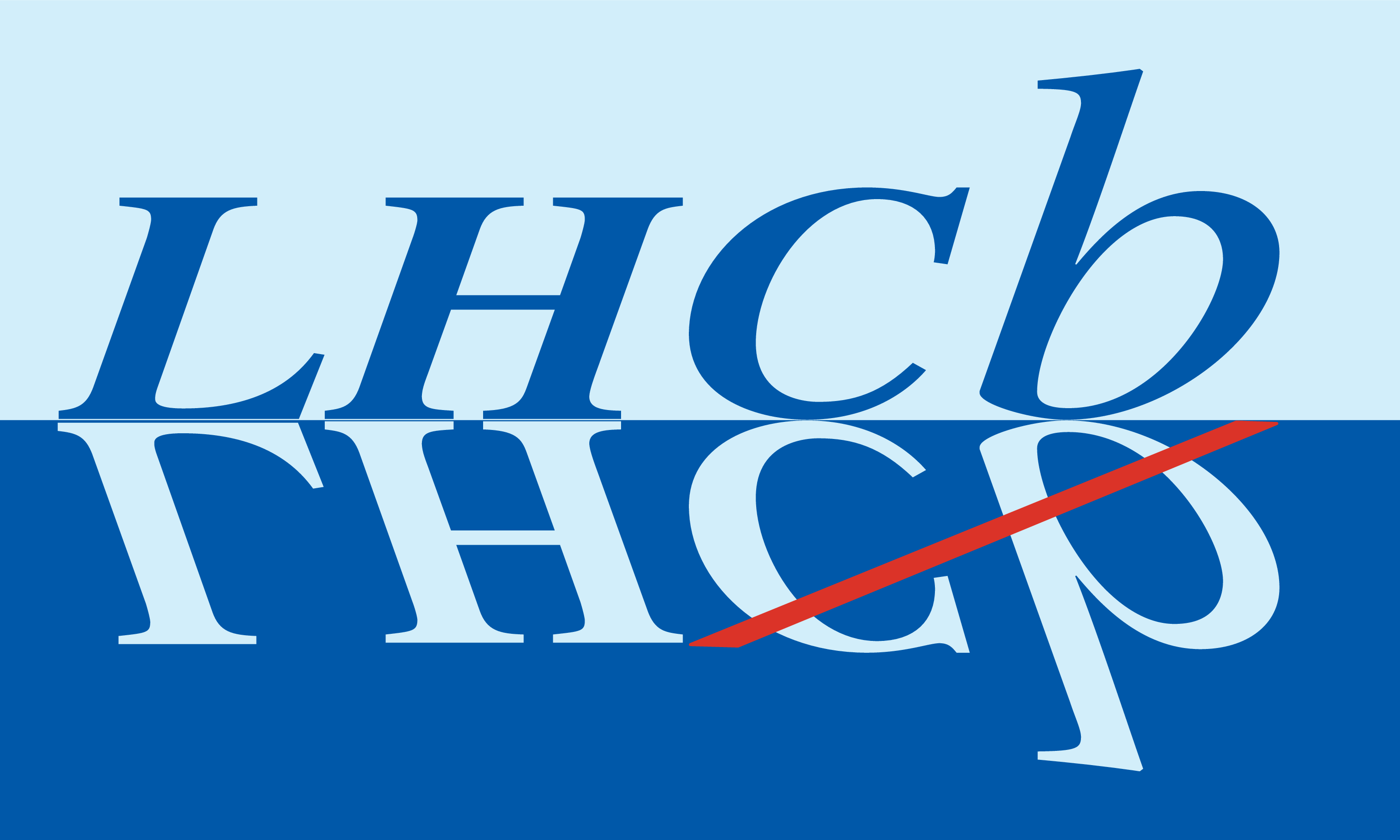}} & &}%
\\
 & & CERN-EP-2022-091 \\  
 & & LHCb-PAPER-2022-006 \\  
 & & \today \\ 
\end{tabular*}

\vspace*{3.8cm}

{\normalfont\bfseries\boldmath\huge
\begin{center}
  \papertitle 
\end{center}
}

\vspace*{1.9cm}

\begin{center}
\paperauthors
\footnote{Authors are listed at the end of this paper.}
\end{center}

\vspace{\fill}

\begin{abstract}
  \noindent
  The interpretation of cosmic antiproton flux measurements from
  space-borne experiments is currently limited by the knowledge of the
  antiproton production cross-section in collisions between primary
  cosmic rays and the interstellar medium. 
  Using collisions of protons with an energy of $6.5$\tev incident on
  helium nuclei at rest in the proximity of the interaction region of
  the \lhcb experiment, the ratio of antiprotons originating from
  antihyperon decays to prompt production is measured for
  antiproton momenta between $12$ and $110\gevc$. 
  The dominant antihyperon contribution, namely \Lbardecay decays from promptly produced \Lbar particles, is also exclusively measured.
  The results complement the  measurement of prompt
  antiproton production obtained from the same data sample. At the energy scale of this measurement, the antihyperon contributions to antiproton production are observed to be significantly larger than predictions of commonly used hadronic production models.
\end{abstract}

\vspace*{1.cm}
\vspace{\fill}

\begin{center}
  Published in Eur.~Phys.~J.~\textbf{C83} (2023) 543 
\end{center}

\vspace{\fill}

{\footnotesize 
\centerline{\copyright~\papercopyright. \href{\paperlicenceurl}{\paperlicence}.}}
\vspace*{2mm}

\end{titlepage}


\newpage
\setcounter{page}{2}
\mbox{~}
%
%
%
%


\renewcommand{\thefootnote}{\arabic{footnote}}
\setcounter{footnote}{0}

\cleardoublepage


\pagestyle{plain} 
\setcounter{page}{1}
\pagenumbering{arabic}


\section{Introduction}
\label{sec:Introduction}
In recent years, the space-borne experiments
PAMELA~\cite{pamela_antiprotons} and AMS-02~\cite{ams_results2021} greatly improved measurements of the abundance of the antiproton, \pbar, component in cosmic rays, which is sensitive to a possible dark matter contribution~\cite{Winkler_WIMP,Boudaud_2019,Cuoco_2019}. 
In the 10--100\gev~\pbar energy range, the interpretation of their measurements requires accurate knowledge of the \pbar production cross-sections in the spallation of cosmic rays in the
interstellar medium~\cite{theo_2015}, which is mainly composed of hydrogen and helium.
The \lhcb experiment has the unique ability to study collisions of the LHC beams with fixed gaseous targets, including helium, reaching the $100\gev$ scale for the nucleon-nucleon centre-of-mass energy, \sqsnn, unprecedented for fixed-target experiments~\cite{LHCb-PUB-2018-015}.  
Using a sample of proton-helium (\pHe) collisions collected in 2016, the production of prompt antiprotons directly in the collisions or through decays of excited states was measured by the \lhcb collaboration~\cite{LHCb-PAPER-2018-031}. These results were the first to use a helium target, and, covering an energy scale where significant violation of Feynman scaling~\cite{FeynmanScaling} occurs, contributed to a better modelling of the secondary \pbar cosmic flux~\cite{Winkler_2017,Donato_2018,Boudaud_2019}.

The uncertainties on \pbar production from weak decays still
limit the interpretation of cosmic \pbar data~\cite{Winkler_2017}. The largest of these contributions 
is due to antineutron decays, which cannot be directly observed in \lhcb but can be estimated from the antiproton measurements and the assumption of isospin symmetry. Another significant contribution, which is less constrained theoretically, comes from decays of  antihyperons, \antihyp . Antiprotons produced in this way are referred to as {\it detached} in the following as they can experimentally be distinguished from prompt antiprotons in the \lhcb experiment by the separation between their production vertex and the primary \pHe collision vertex (PV).

This paper reports a determination of the ratio 
\begin{equation}
\Rinc \equiv \dfrac{\sigma(\pHe \to \antihyp X \to \pbar X)}
{\sigma(\pHe \to \pbar\ped{prompt} X) } 
\end{equation}
of detached to prompt antiprotons in \pHe collisions at  $\sqsnn = 110\gev$ with momentum, \ptot, ranging from $12$ to $110$\gevc and transverse momentum, \pt, between $0.4$ and $4$\gevc, where \textit{X} stands for any arbitrary set of unreconstructed particles. 
Two approaches to the measurement, presented in Secs.~\ref{sec:exclusive}~and~\ref{sec:inclusive}, are followed as described below.

The dominant process, namely \Lbardecay with promptly produced \Lbar particles, is measured relying only on the secondary vertex displacement from the PV and on the decay kinematics. 
The ratio 
\begin{equation}
\Rexc \equiv \dfrac{\sigma(\pHe \to \Lbar X \to \pbar \pip X)}
{\sigma(\decay{\pHe}{\pbar\ped{prompt} X}) } 
\end{equation}
is then determined using the prompt production result~\cite{LHCb-PAPER-2018-031}, obtained from
the same dataset. 

In the second approach, an inclusive measurement of detached antiprotons is performed by exploiting the particle identification (PID) capabilities of the \lhcb detector. Prompt and detached \pbar are 
distinguished by the minimum distance of their reconstructed track 
to the PV, the impact parameter~(IP).
As the relative production yields of different antihyperon states
can be predicted from first principles in statistical models~\cite{Becattini:2010sk}, the $\Rexc/\Rinc$ double ratio is expected to be predicted more reliably than the single ratios. The consistency of the two 
complementary approaches to the analysis is thus verified by comparing the double ratio with model predictions.

The available measurements of the \Rexc ratio, though affected by large uncertainties, hint at a significant increase of this ratio for $\sqsnn > 100 \gev$~\cite{Winkler_2017}. The \lhcb fixed-target configuration is capable of exploring the energy scale where the \Rexc enhancement occurs. 
The contribution to \pbar production from charm and beauty hadron decays
is estimated to be three orders of magnitude smaller than the prompt one, using the measured \ccbar cross-section in the
same fixed-target configuration at \lhcb~\cite{LHCb-PAPER-2018-023} and the known charm branching fractions to baryons~\cite{fragm}. This is negligible compared to the accuracy of this measurement.

\section{The \lhcb detector and its fixed-target operation}
\label{sec:Detector}
The \lhcb detector~\cite{LHCb-DP-2008-001,LHCb-DP-2014-002} is a
single-arm forward spectrometer covering the \mbox{pseudorapidity}
range $2<\eta <5$, designed for the study of particles containing
\bquark or \cquark quarks. The detector includes a high-precision
tracking system consisting of a silicon-strip vertex detector (\velo)
surrounding the proton-proton (\pp) interaction
region~\cite{LHCb-DP-2014-001}, a large-area silicon-strip
detector located upstream of a dipole magnet with a bending power of
about 4\Tm, and three stations of silicon-strip
detectors and straw drift tubes~\cite{LHCb-DP-2017-001} placed
downstream of the magnet. The tracking system provides a measurement
of the momentum of charged particles with a relative
uncertainty that varies from 0.5\% at low momentum to 1.0\% at
200\gevc. The IP is measured with a resolution
of $(15+29/\pt)\mum$, where \pt is measured in \gevc. 
Different types of charged hadrons
are distinguished by using information from two ring-imaging Cherenkov (RICH)
detectors~\cite{LHCb-DP-2012-003}, whose acceptance and performance define the \pbar kinematic range accessible to this study. The first RICH detector has an inner acceptance limited to $\eta<4.4$ and is used to identify antiprotons with momenta between $12$ and $60\gevc$. The second RICH detector, whose Cherenkov threshold for protons is $30\gevc$,
covers the range $3<\eta<5$ and is used for antiproton momenta up to $110\gevc$.
The scintillating-pad detector (SPD) of the calorimeter system is also used in this study. The SMOG (System for Measuring Overlap with Gas) system~\cite{BGI_thesis, LHCb-PAPER-2014-047} 
enables the injection of noble gases with pressure of $\mathcal{O}(10^{-7})$\mbar in the beam pipe section crossing  the \velo, allowing \lhcb to be operated as a fixed-target experiment. 
The online event selection 
is performed by a
trigger~\cite{LHCb-DP-2012-004}, which consists of a hardware stage,
requiring any activity in the SPD detector, and a software stage
asking for at least one reconstructed track in the \velo.
To avoid background from \pp collisions, fixed-target events are acquired only when a bunch in the beam pointing toward \lhcb crosses the nominal interaction region without a corresponding colliding bunch in the other beam.

\section{Data sample and simulation}
\label{sec:Data}
This measurement is performed on data specifically collected for
\pbar production studies in May 2016. Helium gas was injected 
when the two beams circulating in the LHC accelerator consisted 
of proton bunches separated by at least 1~\mus, 40 times the nominal value. In this configuration, spurious \pp collisions are suppressed. 
A sample of \pHe collisions with a $6.5$\tev proton-beam energy ($\sqsnn=110.5$\gev) and corresponding to an integrated luminosity of about $0.5\invnb$ was collected~\cite{LHCb-PAPER-2018-031}. In the proton-nucleon centre-of-mass frame, the \lhcb acceptance corresponds to central and backward rapidities $ -2.8 < y^{\ast} < 0.2$.

Selected events are required to have a reconstructed PV within the fiducial region  \mbox{$-700 < z < +100$\mm}, where the $z$ axis is along the beam direction and $z=0$\mm corresponds to the \lhcb nominal collision
point in the central part of the \velo. The fiducial region is chosen 
to achieve a high efficiency for PV reconstruction in fixed-target collisions and a significant probability that antihyperon decays occur within the \velo. 
Antiproton candidates are reconstructed in the full tracking system exploiting the excellent \Lbar invariant-mass resolution and IP determination. The PV position is required to be compatible with the beam profile and events must have fewer than 5 tracks reconstructed in the \velo with negative pseudorapidity. This selection suppresses to a negligible level the background from interactions with material, decays, and particle showers produced in beam-gas collisions occurring upstream of the \velo. A sample of $33.7$ million reconstructed \pHe collisions satisfying these requirements is obtained from the data.

Simulated data samples of \pHe collisions are produced with the
\eposlhc generator~\cite{epos-lhc}.
The interaction of the generated particles with the detector, and its response, are implemented by using the \geant toolkit~\cite{Allison:2006ve,
*Agostinelli:2002hh} as described in Ref.~\cite{LHCb-PROC-2011-006}. 
The collisions are uniformly distributed along $z$ in the range 
$-1000 < z < +300$\mm, wide enough to cover the fiducial region. When estimating efficiencies, a $z$-dependent weight is applied to simulated events to account for the measured gas pressure variation.
 
This study uses a sample of unbiased simulated inelastic collisions
and several \pbar-enriched samples, where the simulation of the
detector response is performed only if the event generated by \eposlhc
contains a suitable \pbar candidate. In the sample used for the 
inclusive analysis, events must include at least one \pbar with $\pt>0.3\gevc$ 
and $1.9 <\eta < 5.4$. In the sample used for the exclusive analysis,
the antiproton must also come from a \Lbar decay occurring within the acceptance of the \velo. To study the cascade baryon contribution, a sample where the \Lbar decay follows from a \Xibarpdecay decay is also simulated.

\section{Exclusive \texorpdfstring{{\boldmath \Rexc}}{RLbar} measurement}
\label{sec:exclusive}
About 70\%~\cite{CRMC} of the detached antiprotons are expected to originate
from decays of promptly produced \Lbar baryons and can be selected in the \lhcb detector by exploiting the detached decay vertex and the invariant-mass resolution. The decay kinematics allow the antiproton to be identified from the charge and the asymmetry of the longitudinal momenta of the final-state particles with respect to the \Lbar flight direction ($p_{ \text{L}\Lbar}$),  
\begin{equation}
\alpha_{\Lbar} \equiv \dfrac{ p_{\text{L} \Lbar}(\pip) -p_{ \text{L}\Lbar}(\pbar)}{p_{ \text{L}\Lbar}(\pip)+p_{ \text{L}\Lbar}(\pbar)},  
\end{equation}
which is always negative for \Lbar decays~\cite{Armenteros}.
Therefore, the RICH detectors are not used in this approach. 
In order to minimise systematic uncertainties in the measurement of \Rexc, the selection follows as much as possible that used for the prompt measurement~\cite{LHCb-PAPER-2018-031}. In particular, the same fiducial volume, where the  PV reconstruction efficiency cancels in the ratio, and the same kinematic region for the \pbar candidate, $12 < \ptot < 110$\gevc and $0.4 < \pt < 4$\gevc, are required. The analysis is performed in intervals of \ptot and \pt. These intervals are aligned with those used in the prompt measurement, except that some are merged to improve the statistical accuracy.

\subsection{Selection and invariant-mass fit}
\begin{table}
\caption{ Selection requirements for \Lbardecay decays. Symbols are defined in the text.}
\label{tab:Lambda0Selection}
\centering
\begin{tabular}{lc}
  \toprule
  \multirow{3}{*}{Detector acceptance} & 
     $2 < \eta (\pbar) < 5$ \\
   & $2 < \eta (\pip) < 5.5$ \\
   & $2 < \eta (\Lbar) < 5.5;~ \pt (\Lbar) > 0.3\gevc$ \\
  \midrule
  \multirow{3}{*}{Decay geometry} & 
  IP(\Lbar) $< 5$\mm; \chisqdoca $<$ 10 \\ 
  & $\log[ \chisqip(\pbar)] > 1$; $\log[\chisqip(\pip)] > 2$\\
  &  $\fip > 1.5;~ \fipchi > 4$   \\
  \midrule
  \KS veto & $\wm < 490$ or  $\wm > 511 \mevcc$ \\
\midrule
Armenteros-Podolanski & $\Big| \Big(\frac{\alpha_{\Lbar}  + 0.69}{0.18}\Big)^{2} + \Big(\frac{p_{\text{T}\Lbar}(\pip)}{100.4 \mevc}\Big)^{2} - 1 \Big| < 0.39 $\\
\bottomrule				
\end{tabular}
\end{table}

The \Lbar decay candidates are reconstructed from two oppositely charged tracks, which comprise segments in the VELO and in the downstream tracking stations, have a good fit quality and are incompatible with being produced at the PV.
The two-track combinations are selected only if their distance of closest approach is compatible with zero
 using a \chisq test (\chisqdoca). Following previous \Lz production studies in \lhcb~\cite{LHCb-PAPER-2011-005}, large discrimination against combinatorial background 
is obtained by combining the IP information of the \Lbar and the final-state
particles into the linear discriminant
\begin{equation}
\fip \equiv \log\left(\frac{\text{IP}(\pbar)}{1\mm}\right) + \log\left(\frac{\text{IP}(\pip)}{1\mm}\right) - \log\left( \frac{\text{IP}(\Lbar)}{1\mm}\right).
\label{eq:fisher}
\end{equation}
 To take into account the uncertainty on their measurements, a second discriminant \fipchi is constructed by replacing IP in Eq.~(\ref{eq:fisher}) with the \chisqip variable, defined as 
the difference in the vertex-fit \chisq of the PV reconstructed with and
without the track(s) under consideration.
To veto \KSdecay decays, the misreconstructed invariant-mass \wm, obtained by assigning the pion mass to both final-state particles, is required to be incompatible with the \KS mass. Finally, a requirement on the Armenteros-Podolanski plane~\cite{Armenteros} 
$\left(\alpha_{\Lbar},\, p_{\text{T}\Lbar}(\pip) \right)$, where $ p_{\text{T}\Lbar}$ is the transverse momentum with respect to the \Lbar direction, is used.
The selection requirements are listed in Table~\ref{tab:Lambda0Selection}.

\begin{figure}[tb]
    \centering
    \includegraphics[width=0.95\textwidth]{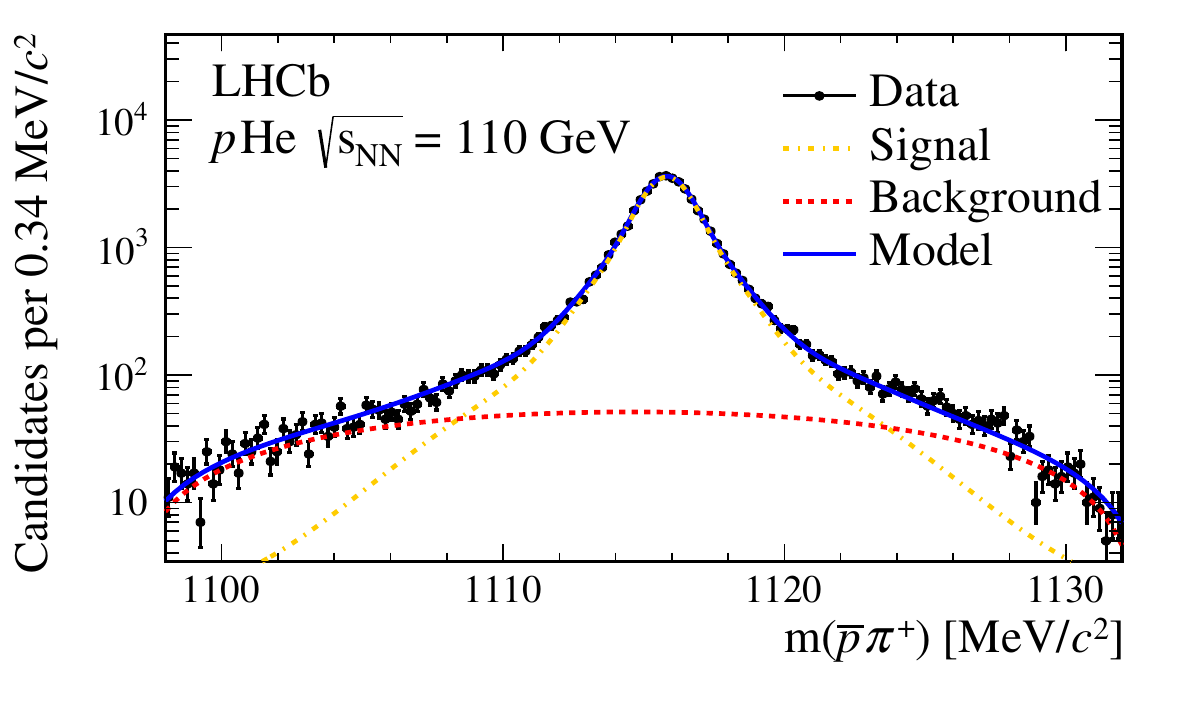}
    \caption{Invariant-mass distribution for the \Lbardecay 
      candidates selected in the \pHe data. The fit model is overlaid on the data.}
    \label{fig:LambdaMass}
\end{figure}

The purity of the selected sample is above $90\%$  in the
unbiased simulation. To subtract the residual background, the invariant-mass distribution of the \Lbar candidates is fitted with the sum of
one Voigtian~\cite{Voigtian} and two Gaussian functions for the signal and a second-order polynomial for the background. This model, validated with simulation, takes into account the bias to the background distribution
from the Armenteros-Podolanski plot requirement and is able to describe the data in all kinematic intervals. The invariant-mass distribution for selected \Lbardecay candidates is shown in Fig.~\ref{fig:LambdaMass} together with a fit integrated over all \ptot and \pt intervals, which results in a yield of $(50.7 \pm 0.3)\cdot 10^3$ \Lbardecay decays.

\subsection{Tracking efficiency}
\label{sec:tracking}
The yields of the selected candidates in each kinematic interval are corrected for the total reconstruction and selection efficiencies. These are
determined as the ratio of signal yields obtained by the invariant-mass fit of the \Lbar candidates in the \pbar-enriched simulated sample to the number of actual candidates generated by the \eposlhc model in the same interval of \ptot and \pt. With this procedure, the efficiency takes into account the resolution effects resulting in migration across kinematic intervals. The largest inefficiency comes from decays occurring downstream of the \velo and can be accurately predicted. For the upstream decays, the average track reconstruction efficiency is determined in simulation to be  $(95.84 \pm 0.04)\%$ for the antiprotons and $(85.40 \pm 0.06)\%$ for the pions, which tend to have a lower momentum. The quoted uncertainties are only due to the finite simulated sample size. These efficiencies are corrected by factors determined from calibration samples in \pp data, which are consistent with unity in all kinematic intervals within their systematic uncertainty of $0.8\%$~\cite{LHCB-DP-2013-002}. 

\begin{figure}[tb]
    \centering
    \includegraphics[width=0.95\textwidth]{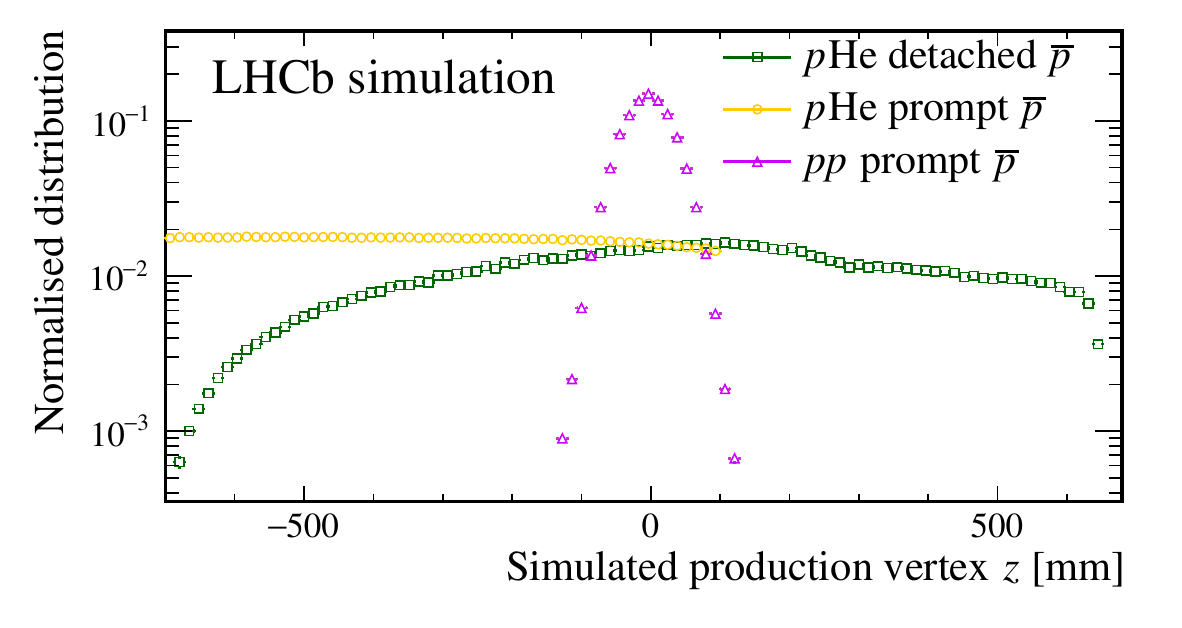}
    \caption{Normalised distributions of the production vertex $z$ coordinate for simulated prompt and detached \pbar in \pHe collisions and for prompt \pbar in simulated \pp collisions in the kinematic range explored in this paper. The PV fiducial region for \pHe collisions is $-700 < z < 100$\mm.} 
    \label{fig:z}
\end{figure}

\begin{figure}[tb]
    \centering
\includegraphics[width = 0.48\textwidth]{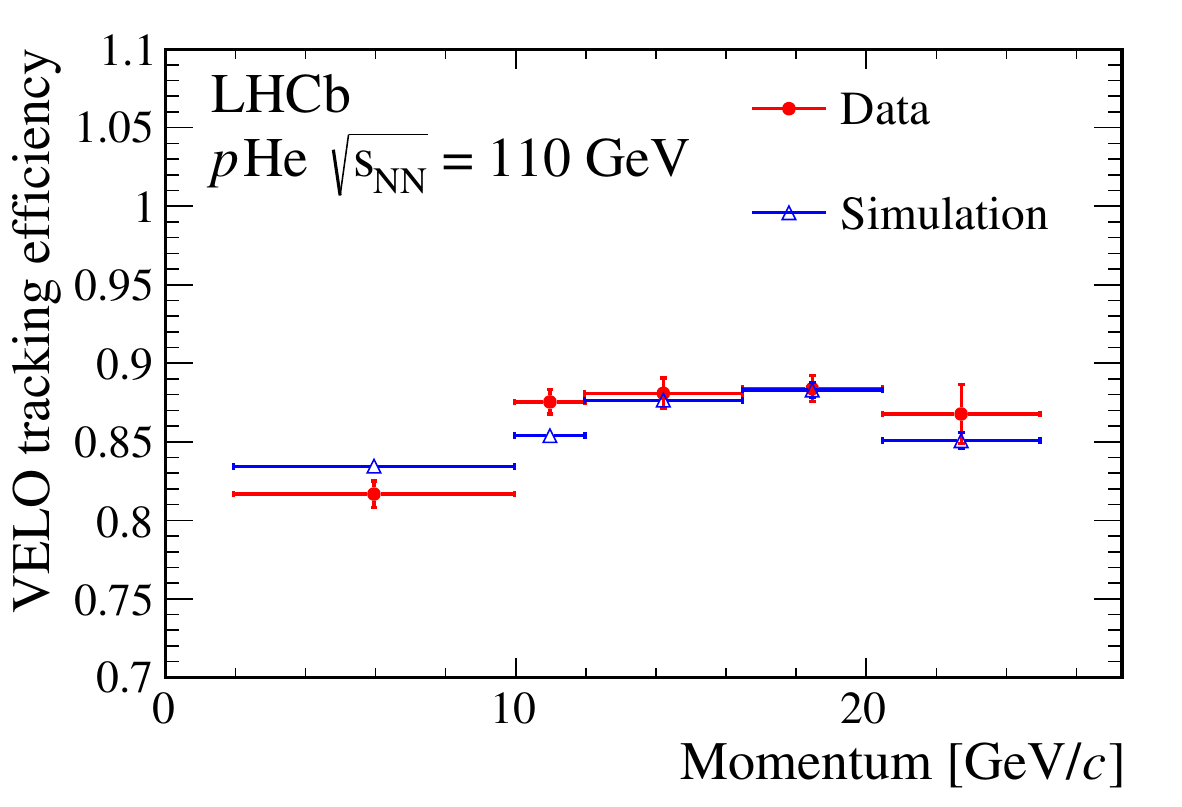} 
\includegraphics[width = 0.48\textwidth]{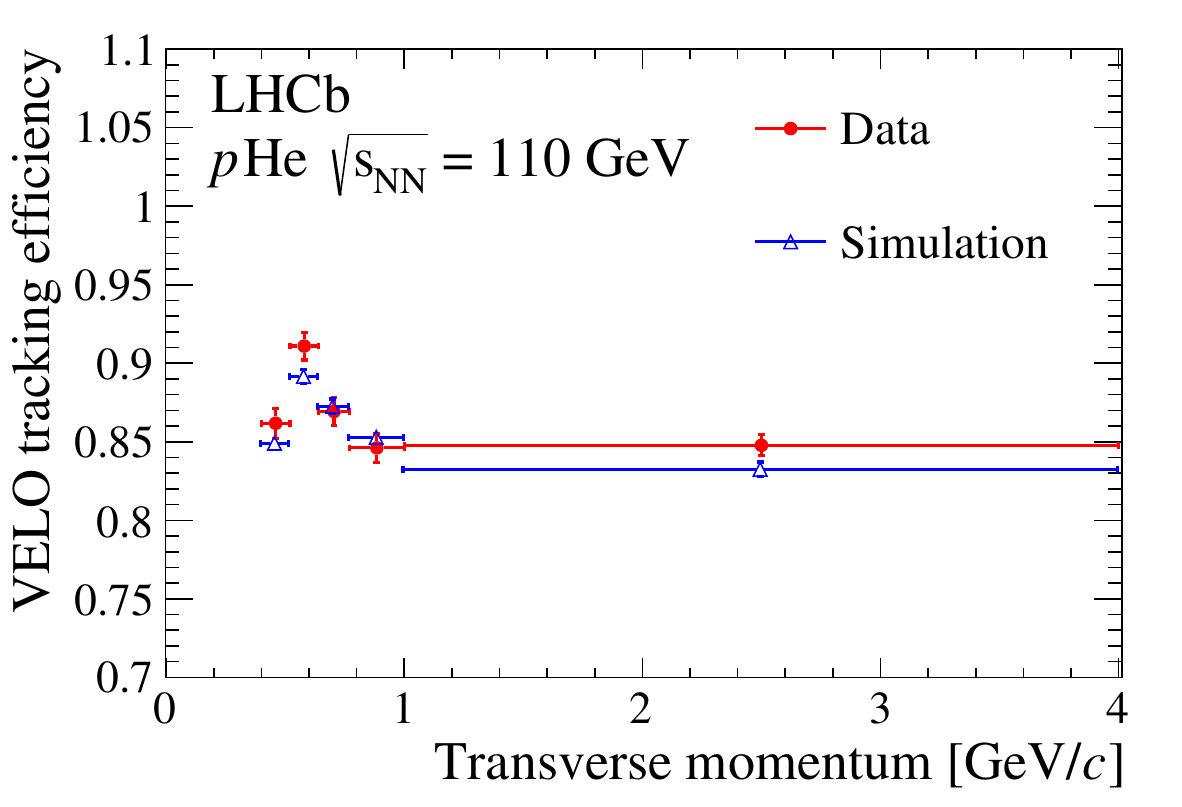} 
\includegraphics[width = 0.48\textwidth]{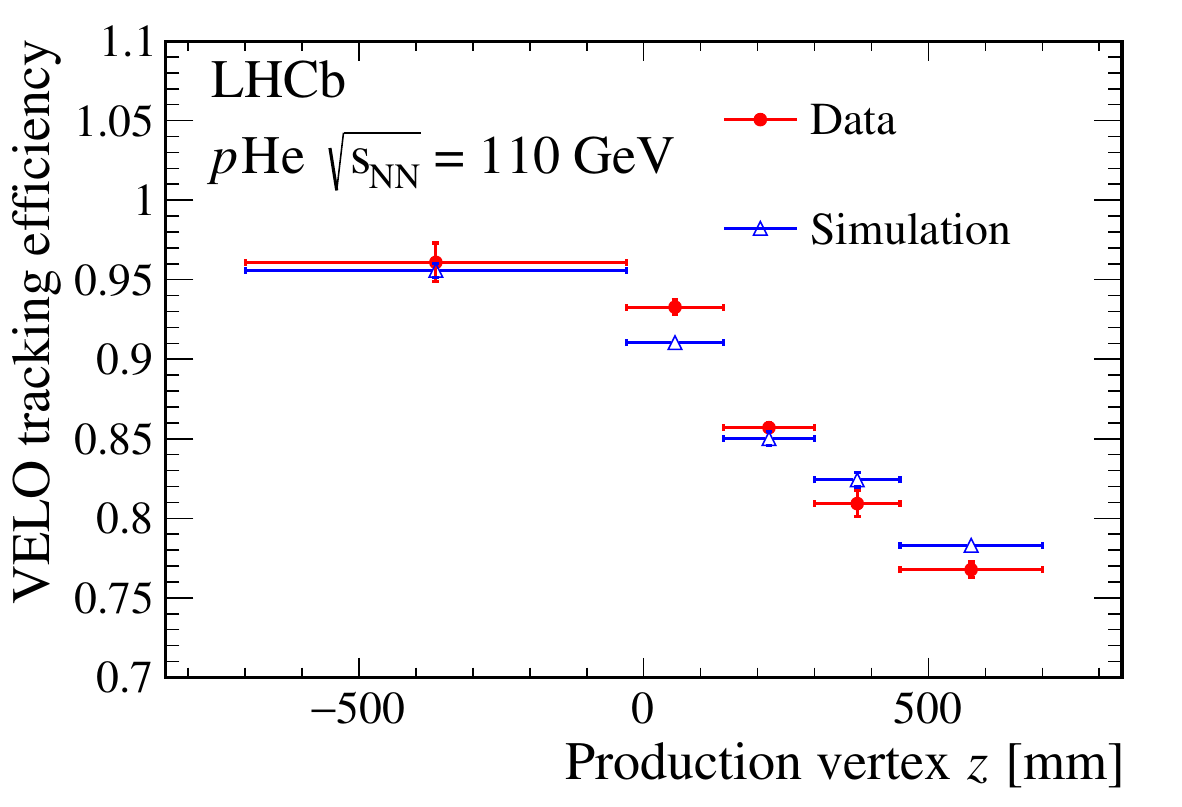} 
\includegraphics[width = 0.48\textwidth]{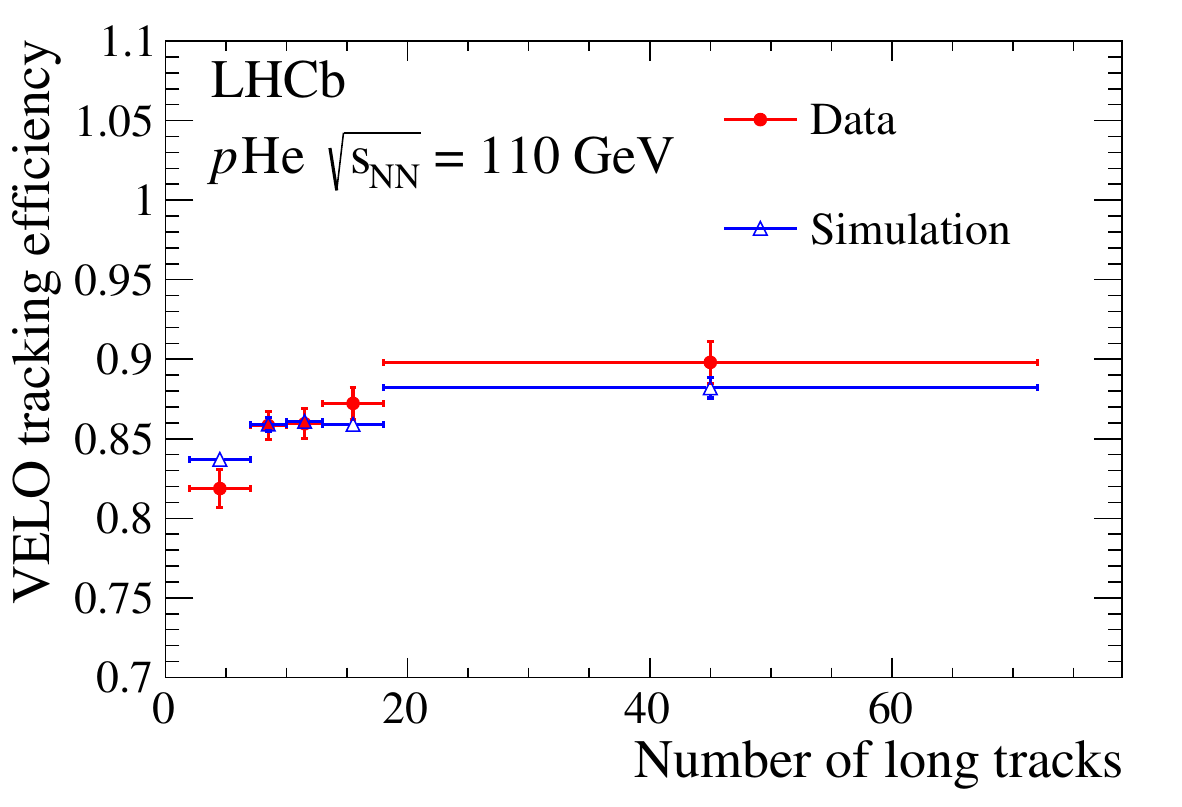} 
    \caption{\velo tracking efficiency for \pbar in \Lbardecay decays as a function of (top left) the particle momentum, (top right) the transverse momentum, (bottom left) the production vertex $z$ coordinate and (bottom right) the number of reconstructed long tracks in the event.}
    \label{fig:VELOeff}
\end{figure}

As illustrated in Fig.~\ref{fig:z}, the tracks considered in this study exhibit a different topology, notably in the \velo, with respect to the prompt tracks from \pp collisions used for calibration, because of the 
larger spread of the fixed-target collision position and the long \Lbar flight distance.
The validation of the \velo tracking efficiency is therefore extended using partially reconstructed \Lbardecay decay candidates in the \pHe collision sample, where the candidate \pbar is reconstructed in the tracking stations upstream and downstream of the magnet but ignoring the information from the \velo. To take into account the degraded resolution of the decay vertex, 
the selection is loosened by requiring $\fip > 1$ and $\fipchi > 3$, while 
the pion, reconstructed using the whole tracking system, is required to be identified by the \rich detectors to compensate for the larger background. The \velo tracking efficiency is estimated from the fraction of candidates in this sample where the partially reconstructed track satisfies the quality requirements for a fully reconstructed track when including the information from the \velo. The \velo efficiencies measured in data and simulation with the same analysis are compared in Fig.~\ref{fig:VELOeff}, notably as a function of the \pbar production vertex position. No significant differences are observed.
The uncertainty on the \pbar reconstruction efficiency in each kinematic interval is expected to cancel in the \Rexc ratio. Therefore, a 0.8\% systematic uncertainty on \Rexc from the \pip tracking efficiency is assigned.

\subsection{Other systematic uncertainties}
A fit model uncertainty is evaluated by repeating the fit with an additional Gaussian component in the 
signal model or with an Argus function~\cite{Argus} for the background model. In both cases, the variation of the result is smaller than the statistical uncertainty in both data and simulation. The online selection requirements are found to be fully efficient in a control sample with randomly selected events. Each of the offline selection requirements listed in Table~\ref{tab:Lambda0Selection} has an efficiency larger than 90\%, with a total selection efficiency of 70\%. The largest inefficiencies are attributed to the requirements on \fip, \fipchi and \wm. The normalised distributions of the two $\mathcal{F}$ discriminants are compared between data and simulated signal. The background contamination is statistically subtracted from the data by modelling the \Lbar invariant-mass distributions and by applying the \sPlot technique~\cite{Pivk:2004ty} with $m(\pbar \pip)$ as discriminating variable. The efficiencies of the requirements in Table~\ref{tab:Lambda0Selection} are measured on the resulting distributions and the difference between data and simulation, amounting to 1\%, is assigned as systematic uncertainty. As a further cross-check of the reliability of the simulation in the wide fiducial region for fixed-target collisions, the analysis is repeated in four equally populated intervals of the PV $z$ position. The efficiency-corrected signal yields are found to agree within the statistical uncertainties.
To check that the \pbar-enriched simulated sample does not bias 
the efficiency estimation, a simulated sample with a looser \Lbar selection is used. The total signal efficiencies are found to agree in all kinematic intervals.

\subsection{Results}
\begin{figure}[tb]
    \centering
    \includegraphics[width = 0.95\textwidth]{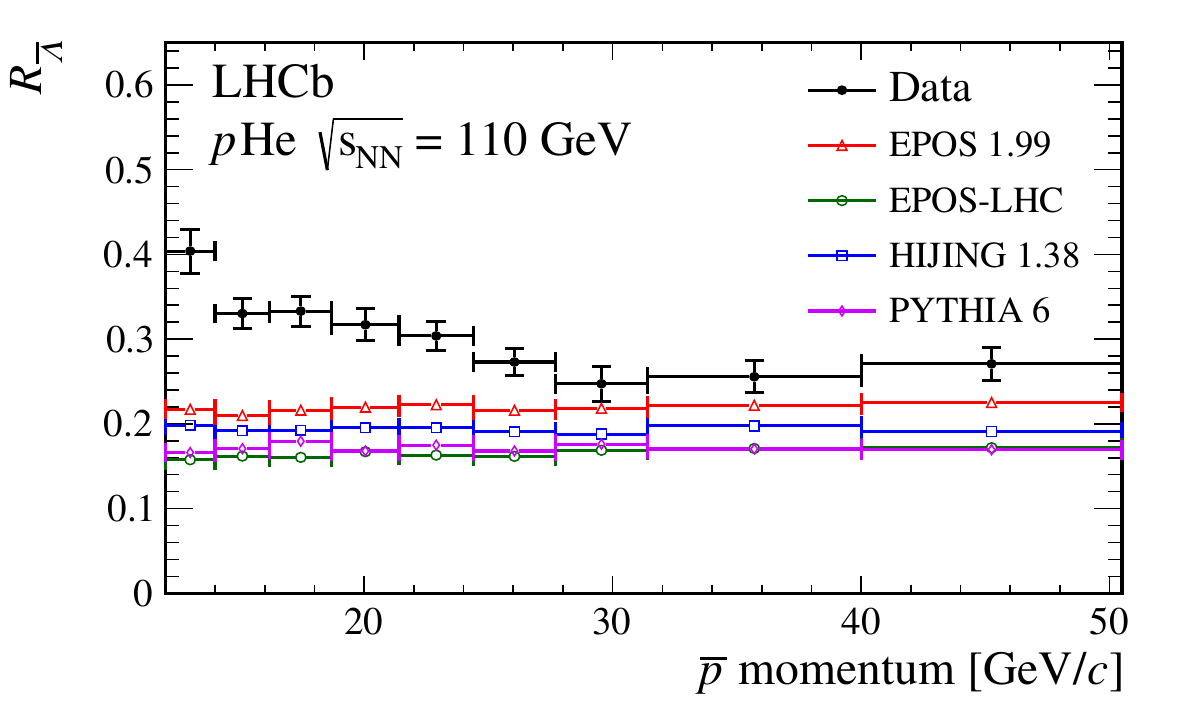}
    \includegraphics[width = 0.95\textwidth]{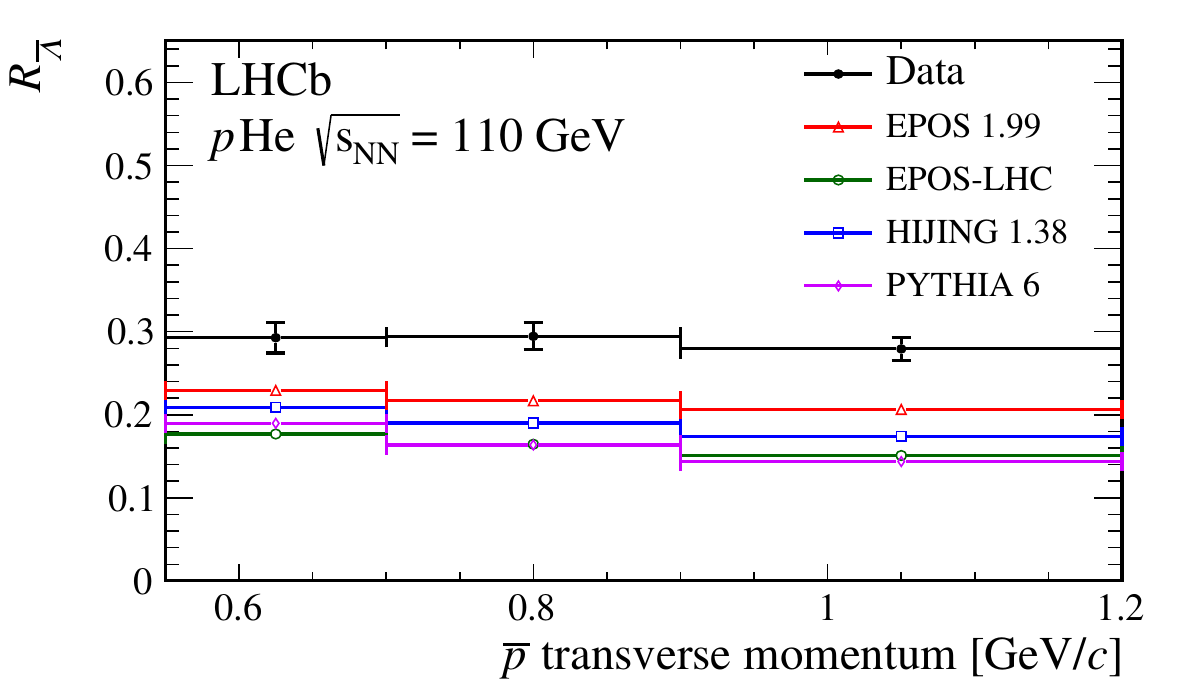}
    \caption{Measured \Rexc as a function of (top) 
      the \pbar momentum for \mbox{$0.55<\pt<1.2$\gevc} and (bottom) the \pbar transverse momentum for \mbox{$12<\ptot<50.5$\gevc}. The measurement is compared to the predictions, in the same kinematic regions, from the \eposold~\cite{EPOS99}, \eposlhc~\cite{epos-lhc}, \hijing~\cite{HIJING} and \pythiaold~\cite{pythia} models, included in the \crmc package~\cite{CRMC}. Error bars on data represent the total uncertainty.} 
    \label{fig:Rexc}
\end{figure}

The ratio \Rexc is determined in each kinematic interval from the measured yield $N_{\Lbar}$ of \Lbardecay decays, the total efficiency  $\epsilon_{\Lbar}$ and the corresponding quantities for prompt \pbar production~\cite{LHCb-PAPER-2018-031} as
\begin{equation}
\Rexc  = \dfrac{N_{\Lbar}}{N\ped{\pbar}} \dfrac{\epsilon\ped{\pbar}}{\epsilon_{\Lbar}}.
\end{equation}
The $N_{\Lbar}$ yields determined from the fits to the \Lbar invariant-mass distributions are corrected by $0.6\%$ to account for the contribution from collisions on the residual gas of the  LHC vacuum contaminating the helium target, as estimated in Ref.~\cite{LHCb-PAPER-2018-031}. The related uncertainty is expected to cancel in the ratio. All significant sources of systematic uncertainty on the \Rexc ratio are listed in Table~\ref{tab:exclSyst}. The leading contributions relate to the particle identification of prompt antiprotons and to the limited size of the produced \Lbar sample. The \Rexc results are illustrated in Fig.~\ref{fig:Rexc2d} and reported in Appendix~\ref{appendixResults} for the kinematic intervals that are common to this and the prompt antiproton production analysis. The results as a function of \ptot (\pt), integrated over the $0.55< \pt < 1.2$\gevc ($12 < \ptot < 50.5$\gevc) region, are shown in Fig.~\ref{fig:Rexc} and compared to widely used hadronic collision models included in the \crmc package~\cite{CRMC}. 
The data indicate that all considered generators significantly underestimate the \Lbar contribution to the \pbar production. 
\begin{table}
	\centering
	\caption{Relative uncertainties on the \Rexc measurement.}
	\label{tab:exclSyst}
	\begin{tabular}{lc}
\toprule
	Particle identification ($N\ped{\pbar}$)       &  $0\% -36\%$ $(<5\%$ for most intervals)\\            
	Statistical uncertainty ($N_{\Lbar}$)          &  $2.2\% -11\%$ $ (<4\%$ for most intervals)\\
	Statistical uncertainty ($N\ped{\pbar}$)       &  $0.5\% -11\%$ $ (<2\%$ for most intervals)\\
	Simulated sample size    ($\epsilon_{\Lbar}$)    &  $1.8\% -4.1\%$ $ (<2\%$ for most intervals)\\
	Simulated sample size    ($\epsilon\ped{\pbar}$) &  $0.4\% -11\% $ $(<2\%$ for most intervals)\\
	Background subtraction ($N\ped{\pbar}$)        & $1.1 \%$  \\
	Selection efficiency ($\epsilon_{\Lbar}$)      & $1\%$     \\
	Tracking efficiency for \pip ($\epsilon_{\Lbar}$)  & $0.8 \%$  \\
\bottomrule
	\end{tabular}
\end{table}
\begin{figure}[tb]
    \centering
    \includegraphics[width =0.95\textwidth]{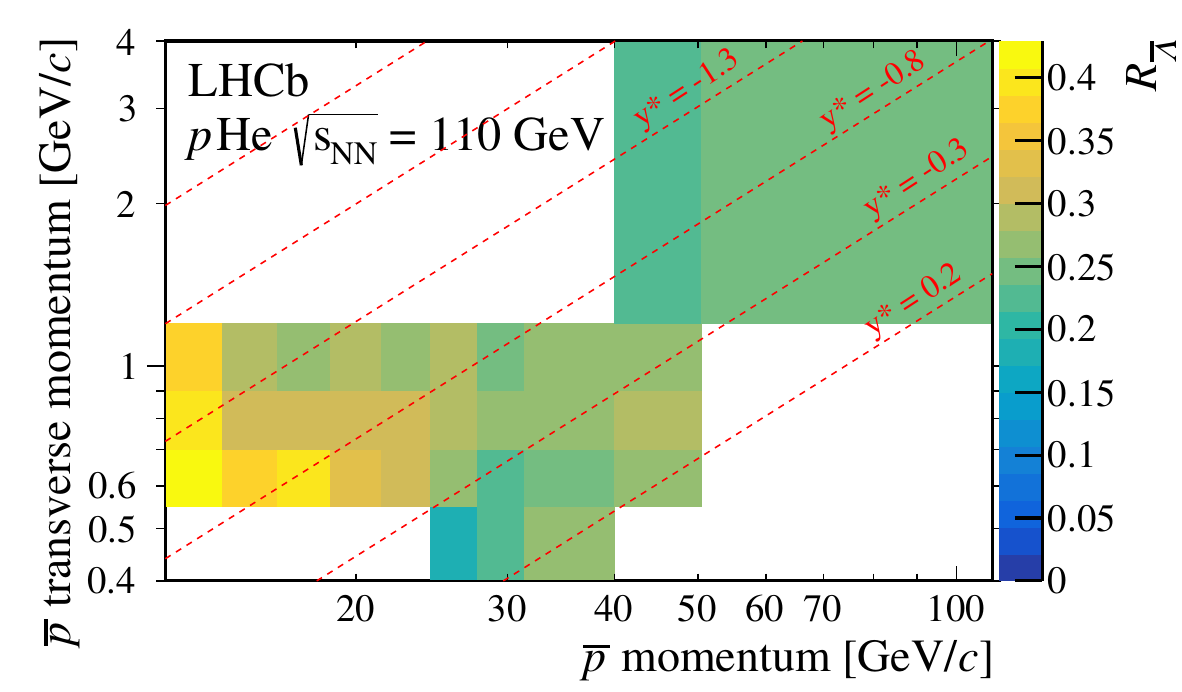}
    \caption{ Measured \Rexc in each of the considered \ptot and \pt intervals.}
    \label{fig:Rexc2d}
\end{figure}

\section{Inclusive \texorpdfstring{{\boldmath \Rinc}}{RHbar} measurement}
\label{sec:inclusive}
\begin{figure}[tb]
    \centering
    \includegraphics[width=0.95\textwidth]{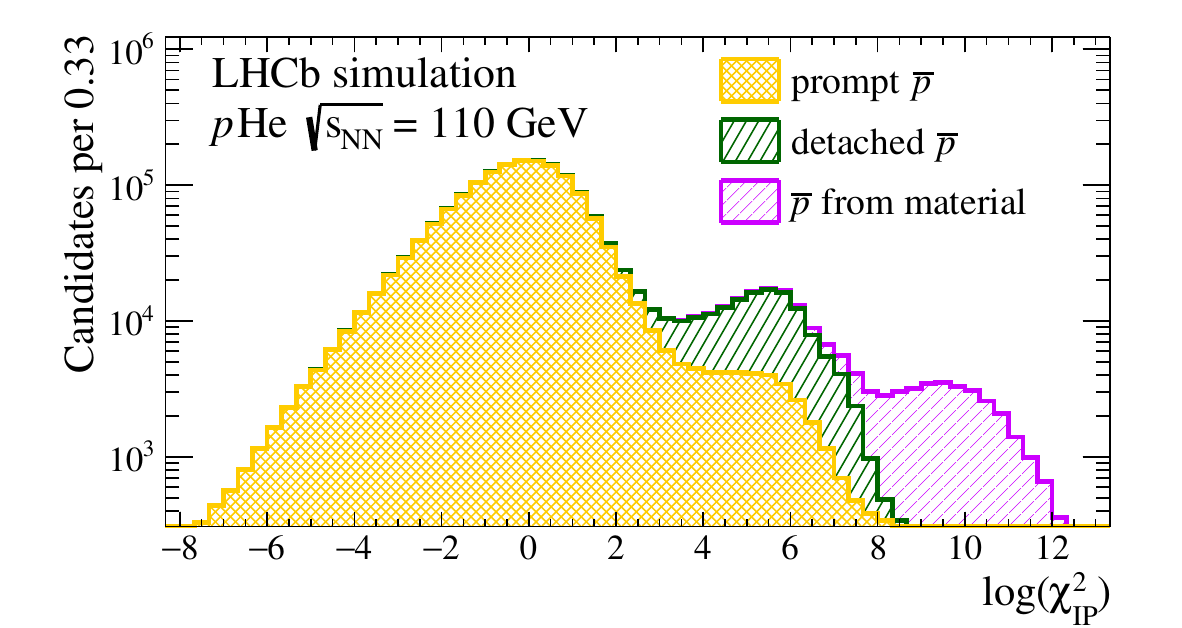}
    \caption{Distributions of the $\log(\chisqip)$ variable for all simulated   
    antiprotons in the \pbar-enriched simulated sample. The  
    contributions from prompt, detached and antiprotons produced in the detector material are separately shown.}
    \label{fig:threepeaks}
\end{figure}

An alternative inclusive approach to the measurement of the detached \pbar yield relies on the PID capabilities of the RICH detectors and on the IP resolution of the \velo, rather than on the reconstruction of the \antihyp decays. In this second analysis, a high-purity \pbar sample is selected through a tight PID requirement. Prompt and detached antiprotons are statistically resolved through a template fit to the distribution of the \chisqip variable. Figure~\ref{fig:threepeaks} shows the $\log(\chisqip)$ distribution for all simulated \pbar in the \pbar-enriched sample. Three contributions can be clearly distinguished, mainly corresponding to prompt, detached, and antiprotons produced in secondary collisions with the detector material. A non-Gaussian tail of the prompt distribution, extending towards the detached \pbar region, is attributed to scattering 
in the material separating the primary \lhc vacuum and the \velo,
as further discussed in Section~\ref{sec:RFfoilSyst}.
Antiproton candidates are selected from negatively charged tracks
reconstructed with a high-quality fit including segments in the \velo and in the tracking stations upstream and downstream of the magnet. The analysis is performed in the (\ptot, \pt) plane, with ten momentum intervals between $12$ and $110$\gevc and five for \pt between $0.4$ and $4\gevc$. The \pbar identification is based on two quantities determined from the response of the RICH detectors: the difference between the logarithm of the likelihood of the proton and pion hypotheses, \dllppi, and that between the proton and kaon hypotheses, \dllpk~\cite{LHCb-DP-2012-003}. A tight selection is used, requiring $\dllppi > 20$ and $\dllpk> 10$, to suppress contamination from misidentified particles. Kinematic intervals at the
boundaries of the RICH capabilities, where the \pbar purity 
predicted in simulation is below 80\%, are removed from the analysis.
The overall predicted purity of the resulting \pbar sample is 97\%.
The numbers of reconstructed prompt and detached
\pbar, $N_{\text{prompt}}$ and $N_{\text{det}}$, are determined from the fit and then corrected for the corresponding efficiencies as estimated
from simulation. These are then used to calculate
\begin{equation}
\Rinc = \dfrac{N_{\text{det}}} { N_{\text{prompt}}} \dfrac{\epsilon_{\text{prompt}}}{\epsilon_{\text{det}}}.
\end{equation}
Efficiencies are determined from the simulation as  the ratio between the number of selected candidates in each interval of reconstructed \ptot and \pt,  and the number generated by the \eposlhc model in the same kinematic interval.

\subsection{Template fit}
\begin{figure}[tb]
    \centering
    \includegraphics[width = 0.95\textwidth]{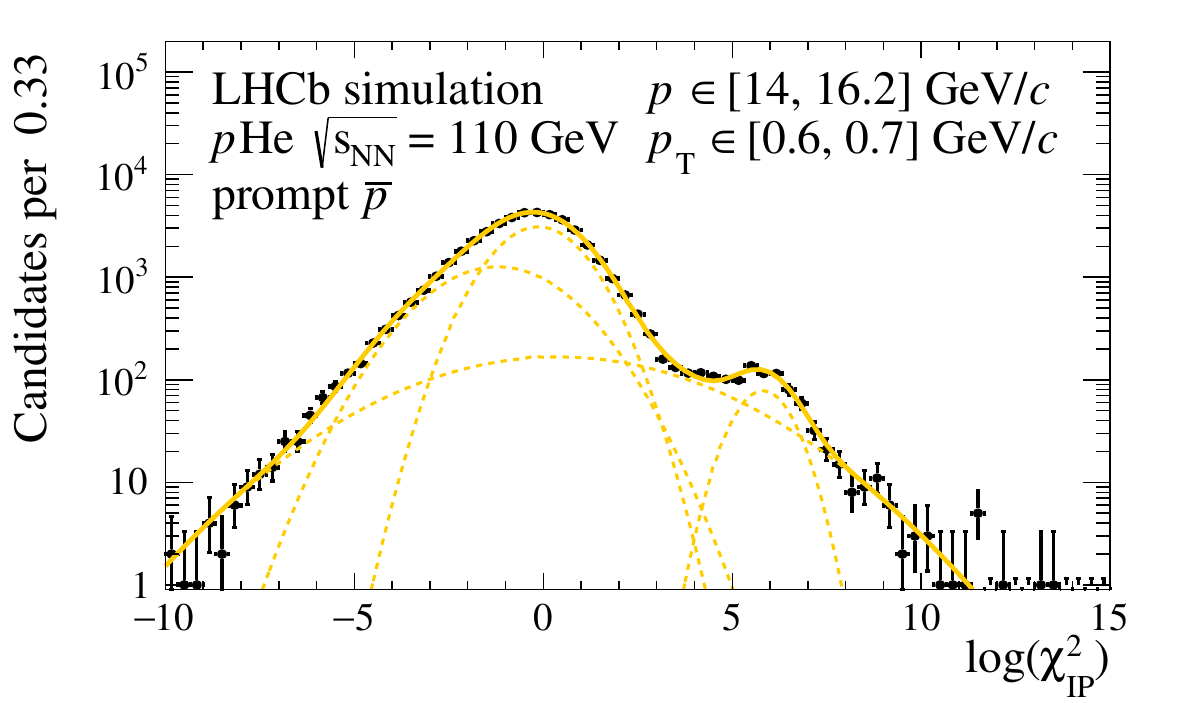}
    \includegraphics[width = 0.95\textwidth]{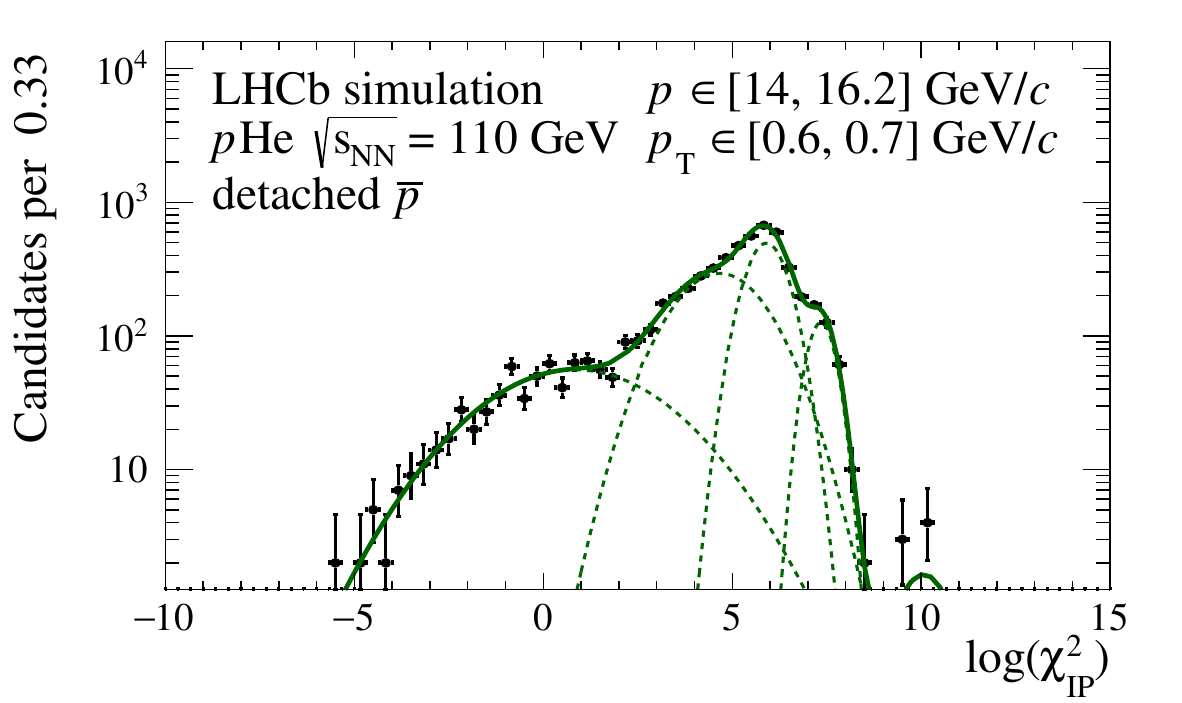} 
    \caption{Distributions of $\log(\chisqip)$ for (top) prompt
      and (bottom) detached antiprotons in the \pbar-enriched
      simulated sample for a kinematic interval in the central region of the considered phase-space. The fit model is overlaid on the data.} 
    \label{fig:templateDrawing}
\end{figure}

Templates for the \chisqip distributions are drawn from the \pbar-enriched simulation in each kinematic interval for different categories of candidates:
three templates for the prompt, detached and secondary \pbar and four templates for misidentified particles, consisting of pions, kaons, electrons and fake tracks. Smoothed curves are obtained through a parametrisation of the probability density functions as a sum of Gaussian functions, whose parameters are obtained from a fit to the simulated event distributions, as illustrated in Fig.~\ref{fig:templateDrawing}. 
For each template the number of Gaussian components, whose parameters
are initialized to random values in the appropriate range, is increased, up to 15, until a good fit is obtained.

\begin{figure}[tb]
    \centering
    \includegraphics[width = 0.95\textwidth]{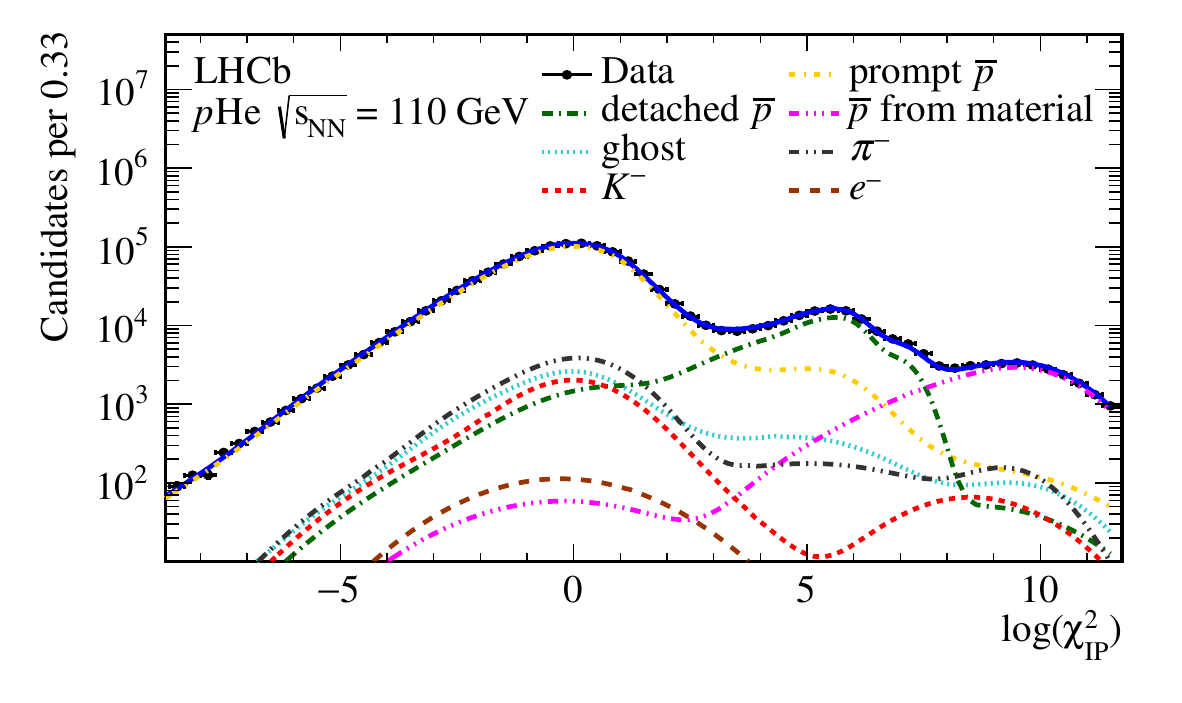}
    \caption{Distributions of the $\log(\chisqip)$ variable in the data sample integrated over all kinematic intervals. The fit model is overlaid on the data.
 }
    \label{fig:fitResult}
  \end{figure}

The template fits are performed with the fractions of the three \pbar components left free to float, while the small contributions from misidentified particles are fixed to the values predicted by the unbiased simulation. The procedure is validated by performing the fit to the unbiased simulated sample and verifying that the obtained abundance for each of the three \pbar categories agrees with the actual value within the statistical uncertainties.
The fit is then applied to the data. Figure~\ref{fig:fitResult}
shows the fit result 
integrated over all kinematic intervals. 
The raw ratio of detached to prompt reconstructed candidates is found to be
$R_{\text{raw}}\equiv N_{\text{det}}/N_{\text{prompt}} = 0.1247 \pm 0.0005$, where the uncertainty is statistical only. This is significantly
larger than the value predicted by the unbiased simulated sample, 
$0.0848 \pm 0.0014$, confirming a sizeable underestimation
of the antihyperon component by the \eposlhc generator.

\subsection{Scattered prompt antiprotons}
\label{sec:RFfoilSyst}
\begin{figure}[tb]
    \centering
    \includegraphics[width = 0.95\textwidth]{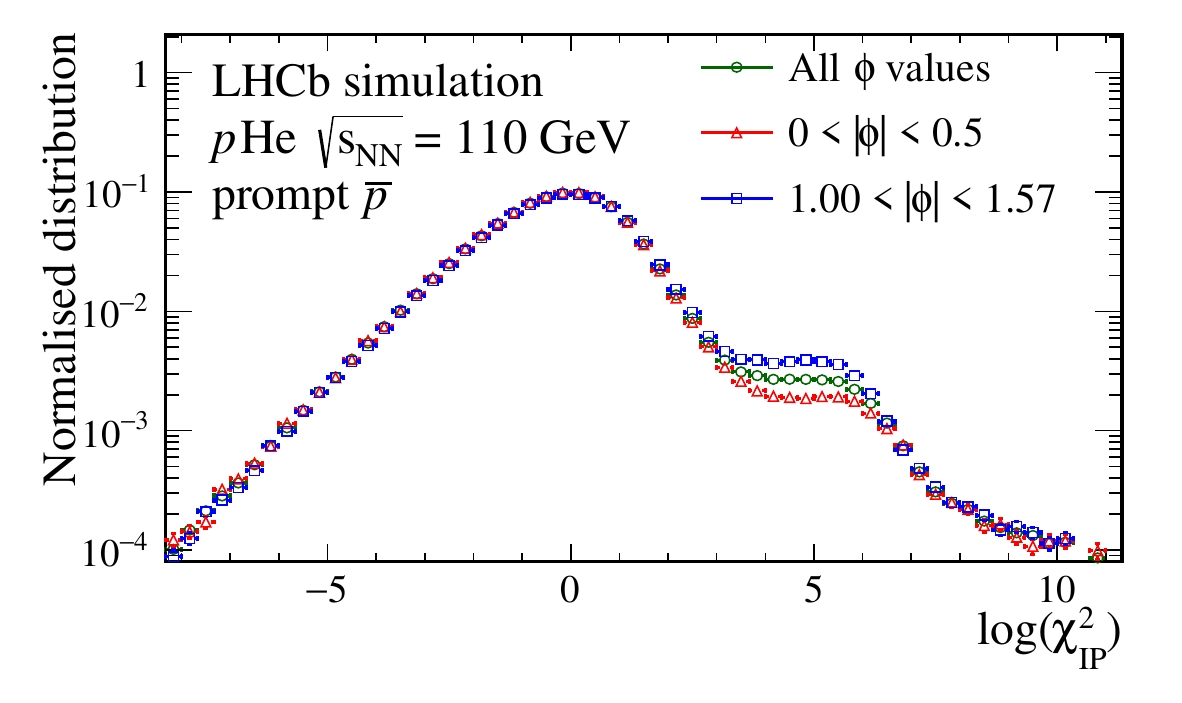}
    \caption{Normalised distributions of the prompt \pbar $\log(\chisqip)$ variable in different ranges of the azimuthal angle $\phi=\text{atan}(p_y/p_x)$.}
    \label{fig:PromptTemplate_Phi}
  \end{figure}

\begin{figure}[tb]
    \centering
    \includegraphics[width = .55\textwidth]{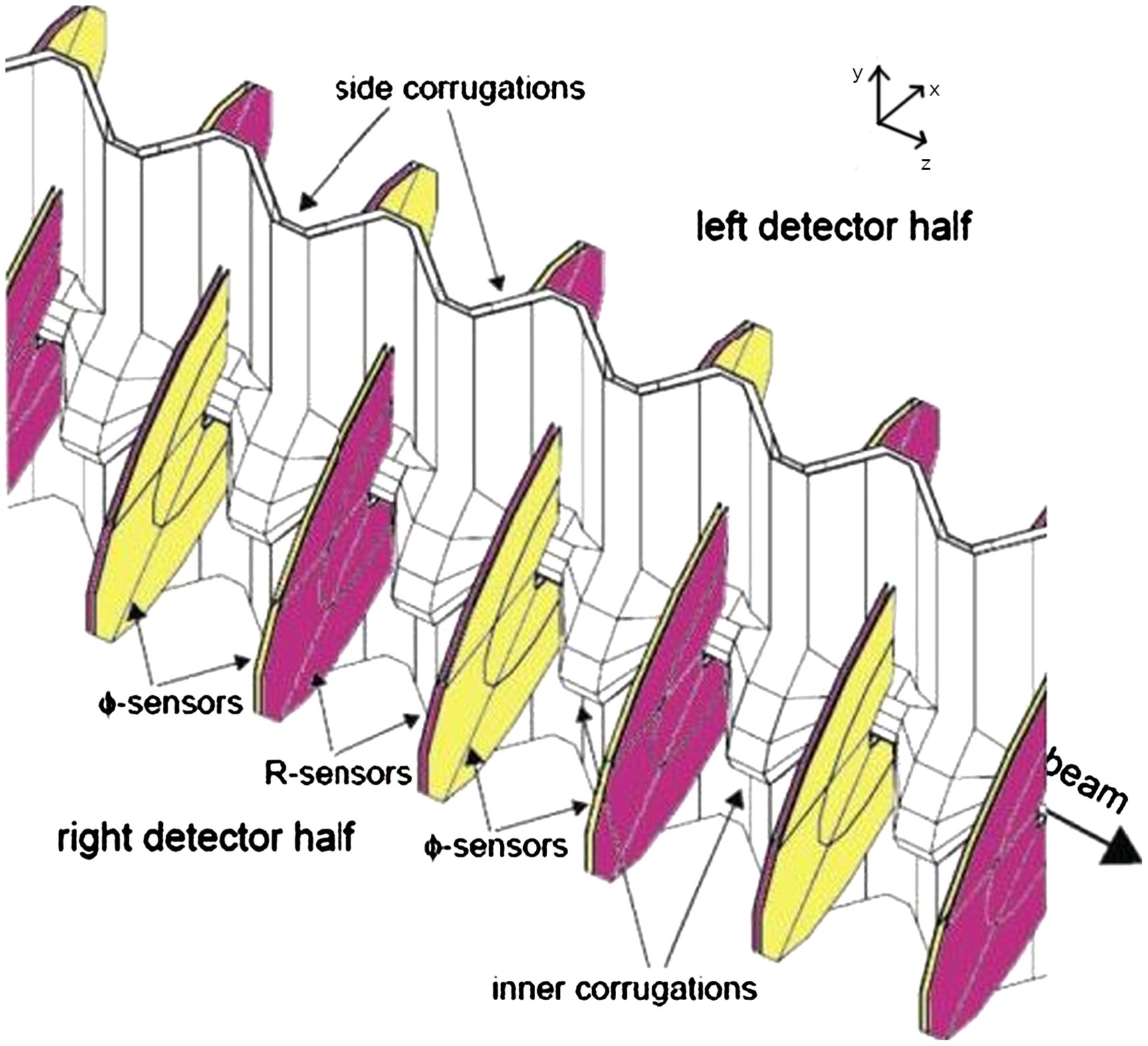}
\caption{Sketch of the \velo~\cite{ALEXANDER2013184}, where the aluminium foils crossed by the particles before entering the \velo volume is visible. The crossed material is maximum for $|\phi| > 1$.}
	\label{fig:VELO_RFfoils}
\end{figure}

\begin{figure}[tb]
    \centering
    \includegraphics[width = .48\textwidth]{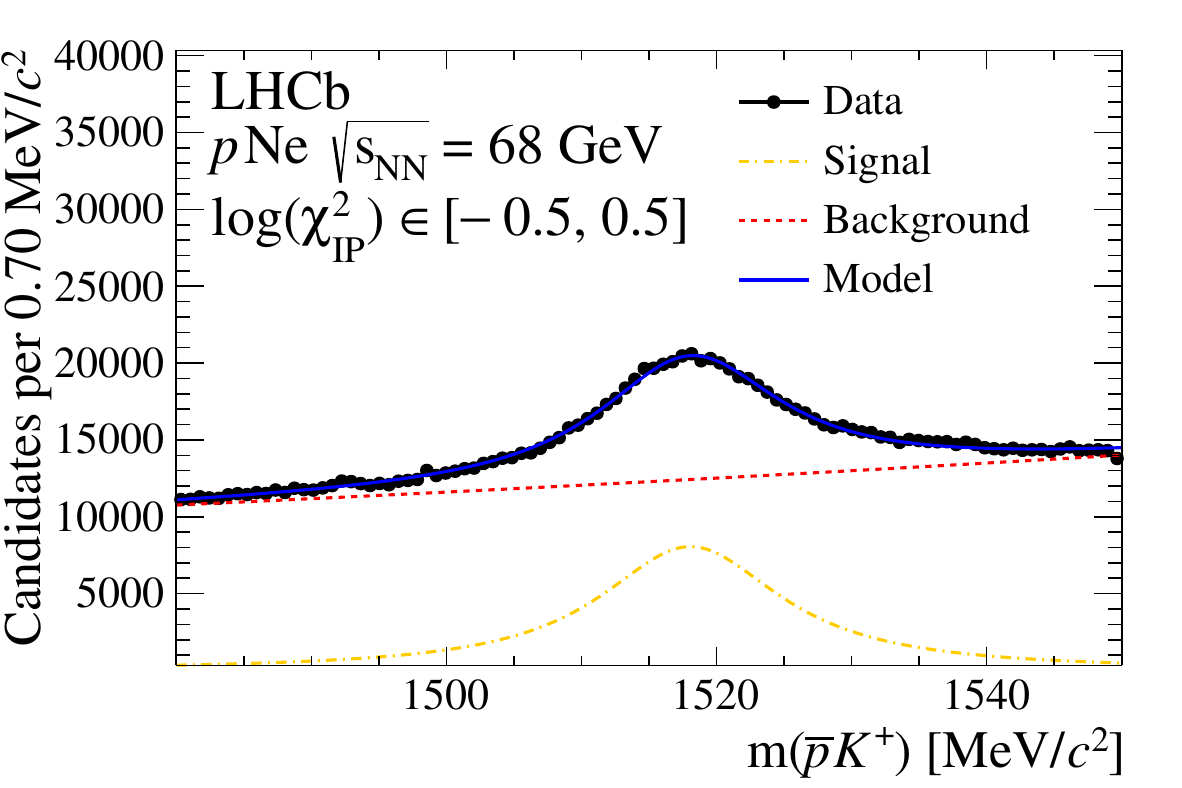}
    \includegraphics[width = .48\textwidth]{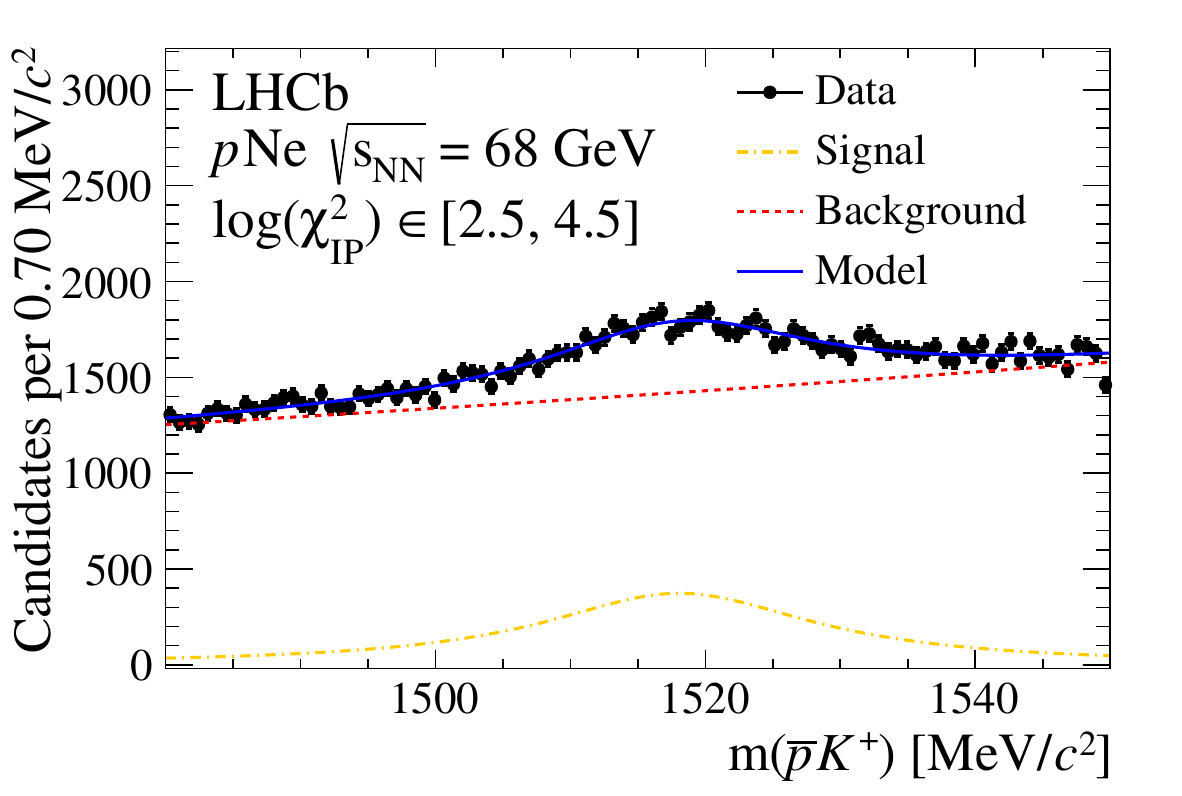}
    \caption{Invariant-mass distributions for the \Lbarstardecay 
      candidates selected in the \pNe data within two intervals of the \pbar $\log(\chisqip)$ variable. The fit model is overlaid on the data.}
    \label{fig:LambdaStar_Fit}
\end{figure}

\begin{figure}[tb]
    \centering
    \includegraphics[width = 0.95\textwidth]{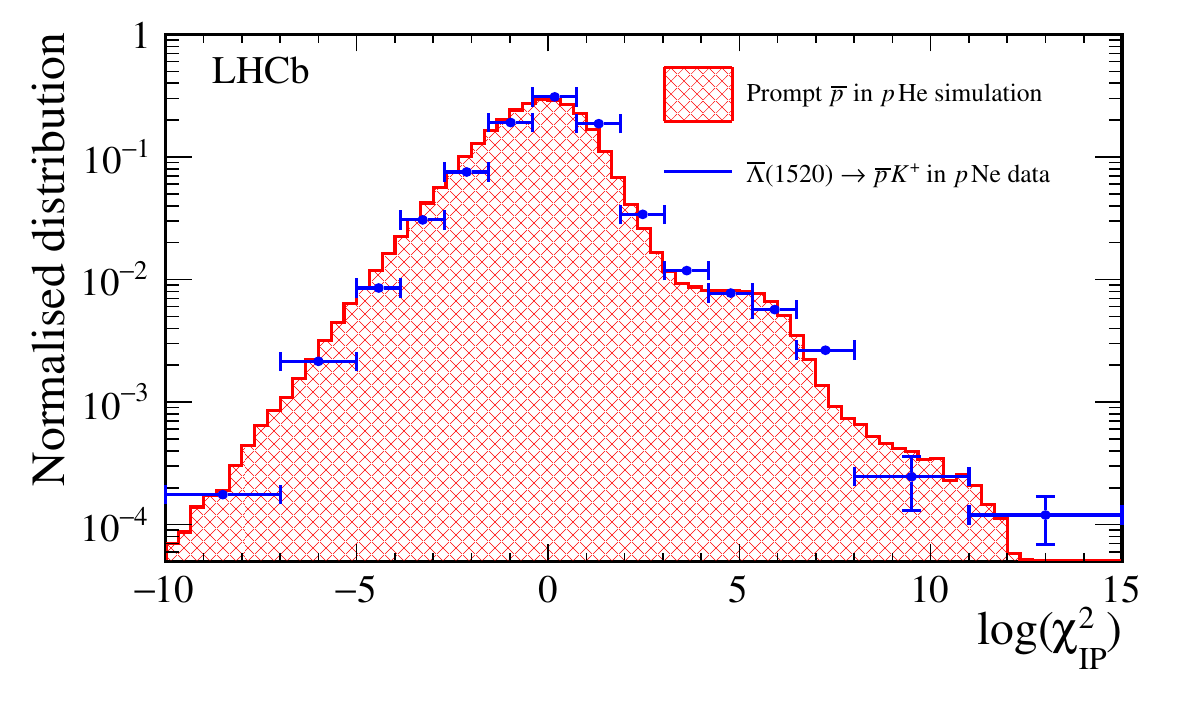}
    \caption{Distributions of the number of signal candidates determined in each interval of the antiproton $\log(\chisqip)$ compared to the prediction of the \pHe simulation for selected prompt antiprotons.}
    \label{fig:LambdaStar_Template}
\end{figure}

Figure~\ref{fig:threepeaks} shows that a significant fraction of simulated prompt \pbar candidates are reconstructed with an IP value well above the expected resolution, compatible with the detached \pbar typical values. As illustrated in Fig.~\ref{fig:PromptTemplate_Phi}, this is due to scattering that may change the track trajectory of the prompt \pbar when the particle crosses the aluminium foil, shown in Fig.~\ref{fig:VELO_RFfoils}, separating the primary \lhc vacuum from the \velo sensor volume.
The tail is indeed found to be strongly dependent on the azimuthal angle $\phi$. The simulation of the material geometry and of the scattering cross-section is therefore critical to the determination of \Rinc. 
A validation using data of the predicted prompt \pbar template is performed by selecting \Lbarstardecay decays. These happen at the primary interaction vertex and, tagged by the invariant-mass of the parent particle and by the kaon identification, provide a prompt \pbar sample in data. Since a small sample size of these decays is selected in the \pHe collision sample, the analysis is performed on the largest fixed-target sample collected during the LHC Run 2, namely a sample of proton-neon~(\pNe) collisions with beam energy of 2.5\tev acquired in 2017. The \pbar candidate is selected with the same requirements as for the inclusive study, while the tagging kaon must satisfy $\dllkpi \equiv \dllppi - \dllpk > 20, \dllpk < 0$
and $\log(\chisqip) < 3$ to enforce prompt decays.
Events are weighted according to the \pbar transverse momentum and
the SPD hit multiplicity to equalize these distributions with those
observed for the prompt \pbar candidates in \pHe data.
The \Lbarstardecay  yield is determined in intervals of the \pbar $\log(\chisqip)$.  
The background is subtracted by fitting the $\pbar\Kp$ 
invariant-mass distribution with a Voigtian function for the signal and an exponential function for the background, as illustrated 
in Fig.~\ref{fig:LambdaStar_Fit}.
The fit parameter representing the signal mass resolution in each interval is independent, as it degrades for increasing values of \chisqip. Figure~\ref{fig:LambdaStar_Template} shows the resulting reconstructed $\log(\chisqip)$ distribution for the prompt \pbar candidates. A reasonable agreement with the template from simulation is found, though differences are expected due to the simplification of the material geometry in the simulation. To estimate the related systematic uncertainty, the inclusive template fit for the sample integrated over all intervals is repeated using this template drawn using data for the prompt \pbar component. The relative variation of $R\ped{\text{raw}}$ is 4.8\% and is assigned as systematic uncertainty.

\subsection{Antihyperon decays}
\label{sec:hyperonCocktail}
The detached \pbar efficiency, in particular the fraction of \antihyp
decays occurring within the \velo, strongly depends on the assumed relative production yields of the different antihyperons contributing to the inclusive \pbar yield. The \eposlhc model predicts that, in the measured kinematic range, 72\% of the \pbar candidates originate from \Lbardecay  decays of promptly produced \Lbar particles, 17\% from \Sbardecay decays, 11\% from \Xibarpdecay and \Xibarzdecay cascade decays, and less than 1\% from \Omegabarp decays. These predictions are expected to be accurate within a relative uncertainty of approximately $20\%$~\cite{Becattini:2010sk}. The assumed values of $\Sigmabarm/\antihyp$ and  $\Xibarp/\Lbar$ ratios are verified with the collision data.

\begin{figure}[tb]
    \centering
    \includegraphics[width = 0.95\textwidth]{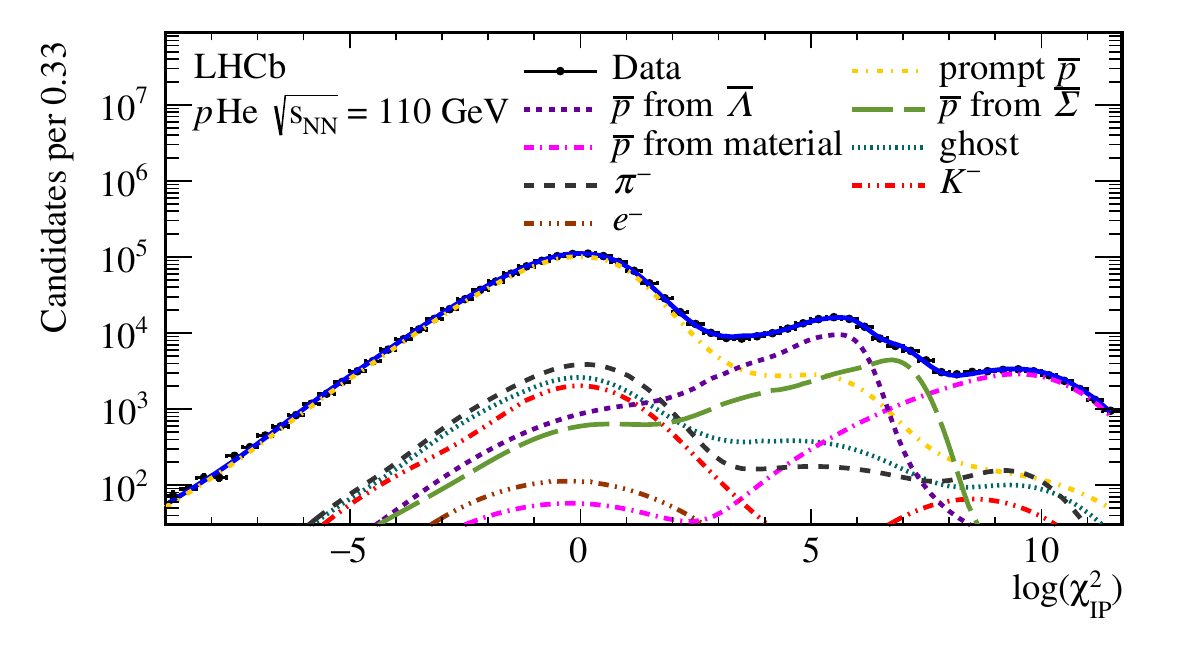}
    \caption{Distribution of the $\log(\chisqip)$ variable in the data sample integrated over all kinematic intervals modelled with independent components for  \Sigmabarm decays (labelled as \pbar from \Sbar) and all other antihyperon decays (\pbar from \Lbar).}
    \label{fig:fitSigma}
\end{figure}

The template fit is expected to have sensitivity to the
contribution of \Sigmabarm decays, as illustrated in Fig.~\ref{fig:fitSigma}. Indeed, when compared to \Lbar decays, the \Sbardecay decay Q-value is larger and antiprotons show on average a larger IP. The fit is repeated with two independent detached
\pbar components: the \Sigmabarm decays and all other decays. The best fit 
fraction of \Sigmabarm is larger than the \eposlhc prediction by a factor $1.13 \pm 0.02$. This correction factor, compatible with the expected theoretical uncertainty, is applied to the simulated sample to recompute the efficiency and correct the detached \pbar templates. While the fit results for $R\ped{raw}$ change less than 1\%, the variation of $\epsilon\ped{det}$ implies a change to \Rinc between $1.2\%$ and $3.8\%$, depending on the kinematic interval. This is assigned as a systematic uncertainty on \Rinc due to the relative \Sigmabarm production.

\begin{figure}[tb]
    \centering
    \includegraphics[width = 0.95\textwidth]{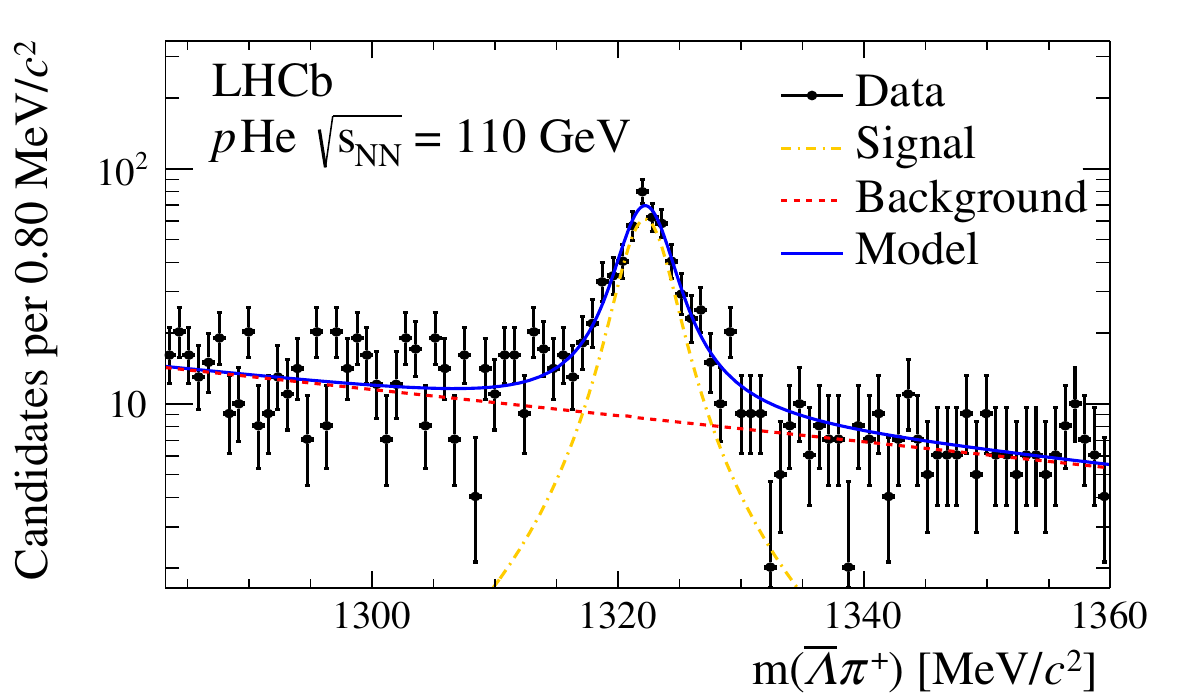}
    \caption{Invariant-mass distribution for the \Xibarpdecay 
      candidates selected in the \pHe data. The fit model is overlaid on the data.}
    \label{fig:Ximass}
\end{figure}

To check the cascade contribution, the \Xibarpdecay yield is directly measured and compared to the \Lbardecay one. Candidates are selected combining a reconstructed \Lbar decay with a \pip candidate track with a distance of closest approach to the  \Lbar trajectory compatible with zero.
Similarly to the prompt \Lbar selection, prompt \Xibarp decays are selected using a linear discriminant: $\log[\chisqip(\Lbar)] +\log[\chisqip(\pip)]-\log[\chisqip(\Xibarp)]>0$. To minimise the systematic bias in the \Xibarp/\Lbar ratio, the final-state \Lbar selection
follows the same requirements as the prompt \Lbar candidates, except that on \chisqip, and a loose selection is chosen for the \pip candidate, without any PID requirement. The $z$ distribution of the decays is also equalised, by weighting the prompt \Lbar candidates to reproduce the observed distribution of \Lbar decay vertices from the reconstructed \Xibarp decays. The invariant-mass distribution of the  \Xibarpdecay  candidates is displayed in
Fig.~\ref{fig:Ximass}, where the fit to determine the signal yield is also shown. The fit model, verified on simulation, uses a Voigtian function for the signal and an exponential function for the background.
The same analysis is performed on the simulated sample and the 
yield ratio $\sigma(\Xibarp)/\sigma(\Lbar)$ is found to be 
larger in data  with respect to the \eposlhc model by a factor $1.09 \pm0.09$.
This factor is used to weight the relative production yield of both 
\Xibarp and \Xibarz baryons in the simulation when deriving the detached \pbar templates in the nominal fits. 
The related uncertainty
corresponds to a systematic uncertainty on \Rinc from cascade production, varying between $0.6\%$ and $0.9\%$, depending on the kinematic interval.

\subsection{Other systematic uncertainties}
\label{sec:inclFitSyst}
The fit model uncertainty   
is estimated by repeating it with the raw binned distributions as templates and is found, in most intervals, to be below 2\%.

Once the probability that antihyperon decays occur within the \velo is taken into account, the reconstruction and selection efficiencies are expected to mostly cancel in the \Rinc ratio. 
Residual differences are still expected from the different distributions
of the production vertex position due to the decaying antihyperons flight distance. The overall tracking efficiency is measured in the simulation as $(94.85 \pm 0.01)\%$ for prompt and $(93.47 \pm 0.05)\%$ for detached antiprotons, with the quoted uncertainties due to the finite simulated sample size. The small difference is mainly due to the lower average number of hits in the \velo for the latter.
As discussed in Section~\ref{sec:tracking}, these geometrical effects are verified to be predicted reliably and no significant systematic uncertainty is assumed. 

A larger bias could be induced by the tight PID selection. In the simulation, its efficiency is measured to be $(64.90 \pm 0.03)\%$ and $(57.74 \pm 0.11)\%$ for prompt and detached antiprotons, respectively. The quoted uncertainties correspond to the finite simulated sample size. These predictions are validated using high-purity \pbar samples from \Lbardecay decays, where the antiproton is identified without using the RICH. A large sample of these  decays is selected from the \pNe collision sample acquired in 2017. A machine-learning-based approach, documented in Ref.~\cite{PID4SMOG}, is used to model the PID response as a function of 12 variables related to the particle trajectory, its reconstruction quality and the event occupancy. This model, applied to the simulated events, is able to reproduce the predicted PID efficiency for the two \pbar categories within the statistical uncertainties, which are lower than 1\%. This demonstrates that the predicted difference is due to geometrical effects and that the RICH response for a given track topology and detector occupancy is accurately simulated. On the other hand, the RICH detector response is affected by low-energy background
that is not accurately simulated. For the selected \Lbardecay  decays in the \pHe sample, the distributions of RICH hit multiplicities differ from those predicted in simulation, and the efficiency of the PID requirement is found  to be larger by a relative 8\% than the predicted value. To check for the sensitivity of the results to the PID selection thresholds, the PID efficiency correction is recomputed in simulation after loosening the selection to reproduce the efficiency measured in data for the detached component. The \Rinc value changes by 0.9\%, which is assigned as the systematic uncertainty on the PID selection. The systematic uncertainties on the predicted fraction of misidentified particles is also evaluated from this check and its effect on \Rinc is found to be negligible.

A systematic uncertainty on the assumed longitudinal profile of the gas target density is assigned from the change of the $\varepsilon\ped{det}/\varepsilon\ped{prompt}$ ratio when introducing the weights to equalize the PV $z$ distribution in data and simulation. It amounts to less than 0.5\% in most kinematic intervals.

\subsection{Results}
\begin{table}
	\centering
	\caption{Relative uncertainties on  the \Rinc measurement. }
	\label{tab:inclSyst}
	\begin{tabular}{lc} \hline
			Prompt \pbar template       & $4.8 \%$    \\
			Statistical uncertainty     & $1.9\% -6.2\%$ $(<2.5\%$ for most intervals)  \\
 			Template parametrisation    & $0 - 5.3\%$ $(<2\%$ for most intervals)   \\ 
			Simulated sample size         &  $1.5\% - 5.8\%$ $(<1.8\%$ for most intervals)  \\
			Production of \Sbar     & $1.2 - 3.8 \%$        \\
			Particle identification     & $0.9 \%$              \\
			Production of \Xiresbar  & $0.6 - 0.9 \%$        \\
			Gas $z$ profile simulation  &  $0.1 - 1.5\%$ $(<0.5\%$ for most intervals)  \\
\hline
	\end{tabular}
\end{table}
\begin{figure}[tb]
    \centering
    \includegraphics[width =0.95 \textwidth]{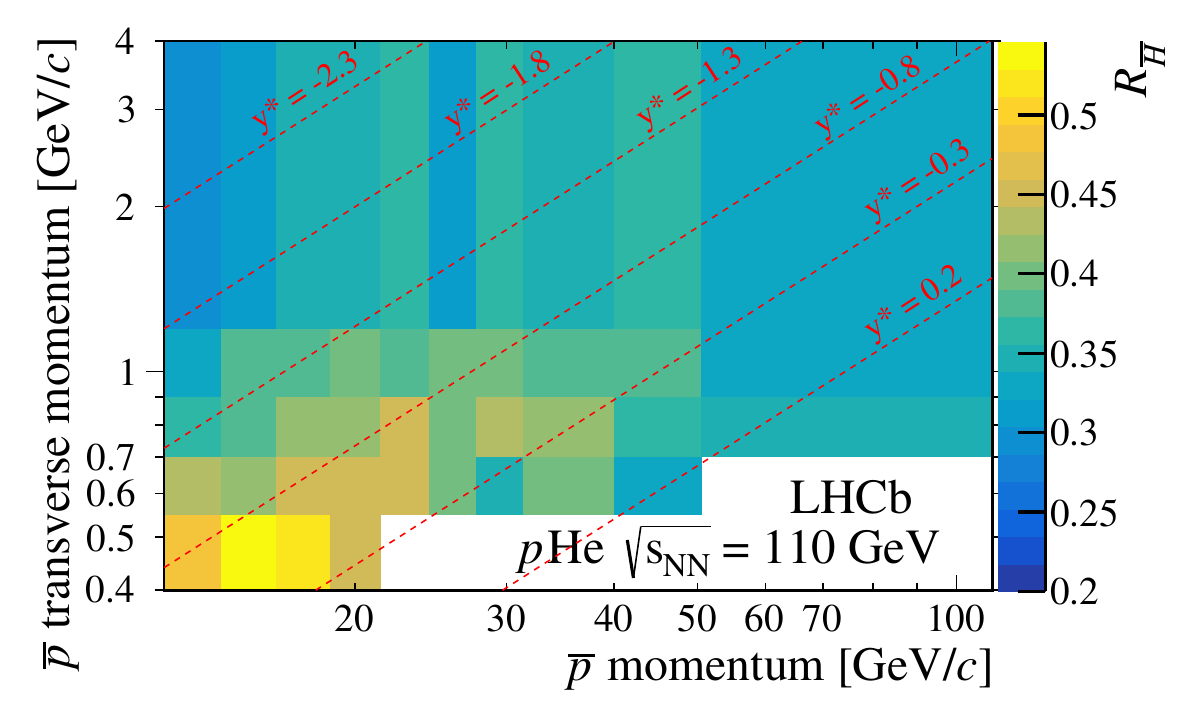}
    \caption{Measured \Rinc in each of the considered \ptot and \pt intervals.}
    \label{fig:Rinc2d}
\end{figure}

\begin{figure}[tb]
    \centering
    \includegraphics[width = .95\textwidth]{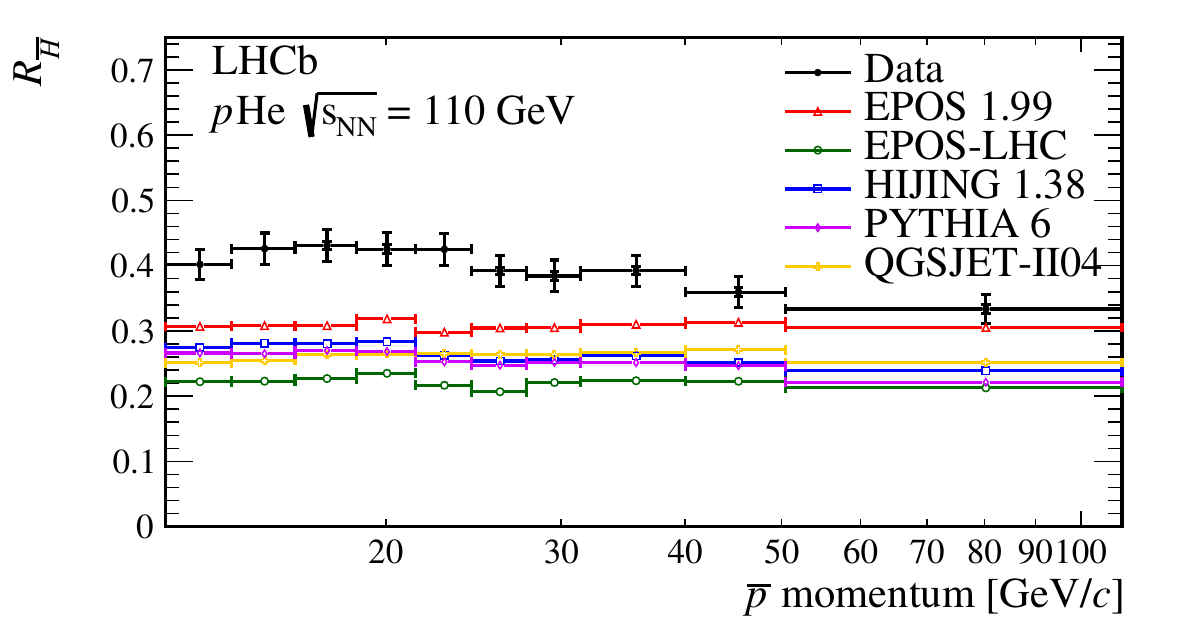}
    \includegraphics[width = .95\textwidth]{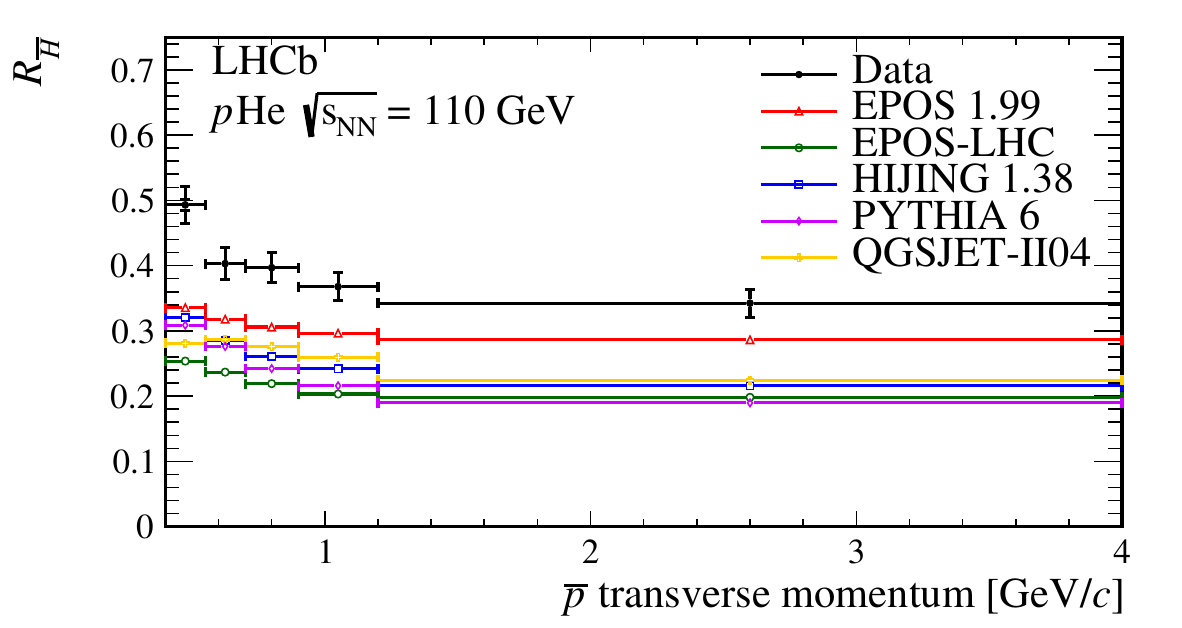}
    \caption{Measured \Rinc as a function of (top) the \pbar momentum for \mbox{$0.4<\pt<4\gevc$} and (bottom) the \pbar transverse momentum for \mbox{$12<\ptot<110\gevc$}. The measurement is compared to predictions, in the same kinematic regions, from the \eposold~\cite{EPOS99}, \eposlhc~\cite{epos-lhc}, \hijing~\cite{HIJING},  \pythiaold~\cite{pythia} and \qgsjet~\cite{QGSJET-II} models, included in the \crmc package~\cite{CRMC}. Error bars on data represent the total uncertainty.} 
    \label{fig:Rinc}
\end{figure}

The results for each kinematic
interval are illustrated in
Fig.~\ref{fig:Rinc2d} and the numerical values are provided in Appendix~\ref{appendixResults}. Table~\ref{tab:inclSyst} summarises the
uncertainties in the \Rinc measurement.  The inclusive results as a function of \ptot or \pt, integrated in the other variable, are shown in Fig.~\ref{fig:Rinc}. As already observed in the \Rexc measurement, the most commonly used hadronic collision generators are shown to underestimate the antihyperon contribution to \pbar production at $\sqsnn=110\gev$.

\label{sec:results}
\begin{figure}[ptb]
    \centering
    \includegraphics[width=0.95\textwidth]{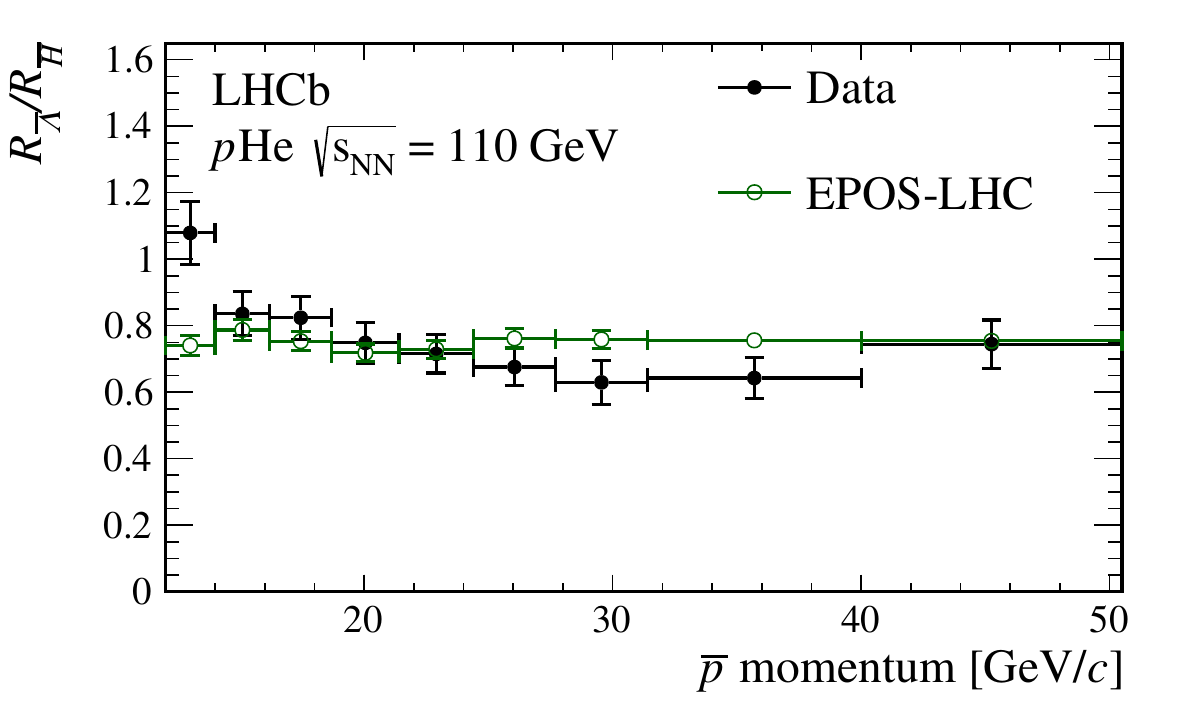}
    \includegraphics[width=0.95\textwidth]{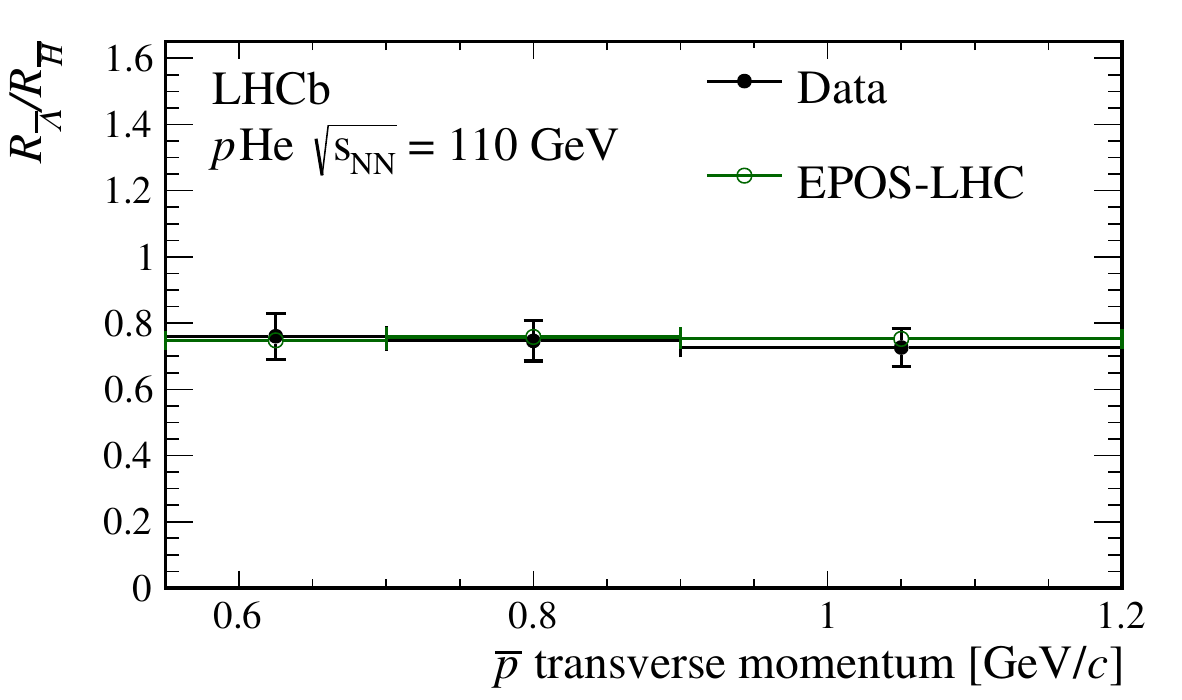}
    \caption{Fraction of antiprotons from decays of promptly produced \Lbar particles to the total yield of detached antiprotons as a function of (top) their momentum for $0.55<\pt<1.2\gevc$ and (bottom) their transverse momentum for $12<\ptot<50.5\gevc$. The data are compared to the \eposlhc~\cite{epos-lhc} prediction for this quantity. Error bars on data represent the total uncertainty.}
    \label{fig:Rratio}
\end{figure}

The ratio $\Rexc/\Rinc$ measured with the inclusive and exclusive approaches is compared with the \eposlhc prediction in Fig.~\ref{fig:Rratio}. As this ratio is predicted more reliably than the inclusive detached \pbar yield, the good agreement between the measured and predicted values provides a mutual validation for the results of the two complementary approaches followed in this paper.

\section{Conclusions}
In conclusion, the production of antiprotons from antihyperon decays relative to the prompt \pbar production
is measured in the
fixed-target configuration of the \lhcb experiment from \pHe
collisions at $\sqsnn = 110\gev$. 
The results confirm previous findings from colliders~\cite{STAR:2006nmo,ALICE:2010vtz,CMS:2011jlm} for an increased \antihyp contribution with
respect to the $\sqsnn~\sim~10\gev$ scale probed in past fixed-target experiments and indicate a sizeable underestimation of this contribution in most hadronic production models used in cosmic ray physics. 
A significant dependence of \Rinc on the \pbar momentum is observed. This
effect is not usually considered in the modelling
of the secondary \pbar component in cosmic rays, where \Rinc is assumed to depend only on \sqsnn~\cite{Winkler_2017,Donato_2018}.
These results are thus expected to provide a valuable input to improve the predictions for the secondary \pbar cosmic flux.
\clearpage

\section*{Acknowledgements}
%
%
\noindent We express our gratitude to our colleagues in the CERN
accelerator departments for the excellent performance of the LHC. We
thank the technical and administrative staff at the LHCb
institutes.
We acknowledge support from CERN and from the national agencies:
CAPES, CNPq, FAPERJ and FINEP (Brazil); 
MOST and NSFC (China); 
CNRS/IN2P3 (France); 
BMBF, DFG and MPG (Germany); 
INFN (Italy); 
NWO (Netherlands); 
MNiSW and NCN (Poland); 
MEN/IFA (Romania); 
MICINN (Spain); 
SNSF and SER (Switzerland); 
NASU (Ukraine); 
STFC (United Kingdom); 
DOE NP and NSF (USA).
We acknowledge the computing resources that are provided by CERN, IN2P3
(France), KIT and DESY (Germany), INFN (Italy), SURF (Netherlands),
PIC (Spain), GridPP (United Kingdom), 
CSCS (Switzerland), IFIN-HH (Romania), CBPF (Brazil),
Polish WLCG  (Poland) and NERSC (USA).
We are indebted to the communities behind the multiple open-source
software packages on which we depend.
Individual groups or members have received support from
ARC and ARDC (Australia);
Minciencias (Colombia);
AvH Foundation (Germany);
EPLANET, Marie Sk\l{}odowska-Curie Actions and ERC (European Union);
A*MIDEX, ANR, IPhU and Labex P2IO, and R\'{e}gion Auvergne-Rh\^{o}ne-Alpes (France);
Key Research Program of Frontier Sciences of CAS, CAS PIFI, CAS CCEPP, 
Fundamental Research Funds for the Central Universities, 
and Sci. \& Tech. Program of Guangzhou (China);
GVA, XuntaGal, GENCAT and Prog.~Atracci\'on Talento, CM (Spain);
SRC (Sweden);
the Leverhulme Trust, the Royal Society
 and UKRI (United Kingdom).



\newpage
\appendix
\section{Numerical results}
\label{appendixResults}
The numerical results for the measurements of the \Rexc and \Rinc ratios, presented in Secs.~\ref{sec:exclusive} and \ref{sec:inclusive}, are reported in Table~\ref{tab:Results_1} in intervals of the antiproton momentum and transverse momentum.

\begin{table} [ht]
\centering 
\caption{Measured \Rinc and \Rexc ratios in intervals of the antiproton momentum and transverse momentum, measured in \gevc units. The average momentum and transverse momentum, as predicted by the \eposlhc generator for prompt antiprotons, are also listed for each interval in \gevc units. The uncertainty is split into an uncorrelated component, denoted with $\delta\ped{uncorr}$, and a component that is fully correlated among the kinematic intervals, denoted $\delta\ped{corr}$.}
\label{tab:Results_1}
\end{table}
\noindent
\begin{tabular}{cccccc|ccc|ccc} 
\toprule 
$p$\ap{min}
 &
$\langle p \rangle$
 &
$p$\ap{max}
 &
$p_{\mathrm{T}}^\text{min}$
 &
$\langle p_{\mathrm{T}} \rangle$
 &
$p_{\mathrm{T}}^\text{max}$
 &
\Rinc
 &
$\delta\ped{uncorr} $
 &
$\delta\ped{corr} $
 &
\Rexc
 &
$\delta\ped{uncorr} $
 &
$\delta\ped{corr} $
\\  \midrule 
12.0
 &
12.98
 &
14.0
 &
0.40
 &
0.47
 &
0.55
 &
0.477
 &
0.013
 &
0.025
 &
---
 &
---
 &
---
\\   
12.0
 &
12.99
 &
14.0
 &
0.55
 &
0.62
 &
0.70
 &
0.433
 &
0.012
 &
0.023
 &
0.429
 &
0.019
 &
0.022
\\   
12.0
 &
12.99
 &
14.0
 &
0.70
 &
0.79
 &
0.90
 &
0.371
 &
0.012
 &
0.020
 &
0.395
 &
0.036
 &
0.019
\\   
12.0
 &
12.99
 &
14.0
 &
0.90
 &
1.02
 &
1.20
 &
0.337
 &
0.016
 &
0.019
 &
0.377
 &
0.042
 &
0.013
\\   
12.0
 &
13.00
 &
14.0
 &
1.20
 &
1.47
 &
4.00
 &
0.303
 &
0.026
 &
0.018
 &
---
 &
---
 &
---
\\   
14.0
 &
15.08
 &
16.2
 &
0.40
 &
0.47
 &
0.55
 &
0.546
 &
0.018
 &
0.029
 &
---
 &
---
 &
---
\\   
14.0
 &
15.09
 &
16.2
 &
0.55
 &
0.62
 &
0.70
 &
0.423
 &
0.011
 &
0.022
 &
0.370
 &
0.017
 &
0.019
\\   
14.0
 &
15.09
 &
16.2
 &
0.70
 &
0.79
 &
0.90
 &
0.387
 &
0.010
 &
0.021
 &
0.317
 &
0.012
 &
0.016
\\   
14.0
 &
15.09
 &
16.2
 &
0.90
 &
1.02
 &
1.20
 &
0.383
 &
0.014
 &
0.022
 &
0.290
 &
0.018
 &
0.011
\\   
14.0
 &
15.10
 &
16.2
 &
1.20
 &
1.48
 &
4.00
 &
0.311
 &
0.022
 &
0.019
 &
---
 &
---
 &
---
\\   
16.2
 &
17.43
 &
18.7
 &
0.40
 &
0.47
 &
0.55
 &
0.518
 &
0.022
 &
0.027
 &
---
 &
---
 &
---
\\   
16.2
 &
17.43
 &
18.7
 &
0.55
 &
0.62
 &
0.70
 &
0.442
 &
0.012
 &
0.024
 &
0.395
 &
0.018
 &
0.019
\\   
16.2
 &
17.43
 &
18.7
 &
0.70
 &
0.79
 &
0.90
 &
0.412
 &
0.010
 &
0.022
 &
0.317
 &
0.012
 &
0.016
\\   
16.2
 &
17.44
 &
18.7
 &
0.90
 &
1.02
 &
1.20
 &
0.374
 &
0.012
 &
0.021
 &
0.273
 &
0.015
 &
0.011
\\   
16.2
 &
17.44
 &
18.7
 &
1.20
 &
1.49
 &
4.00
 &
0.352
 &
0.020
 &
0.020
 &
---
 &
---
 &
---
\\   
18.7
 &
20.03
 &
21.4
 &
0.40
 &
0.47
 &
0.55
 &
0.447
 &
0.024
 &
0.025
 &
---
 &
---
 &
---
\\   
18.7
 &
20.03
 &
21.4
 &
0.55
 &
0.62
 &
0.70
 &
0.458
 &
0.013
 &
0.024
 &
0.342
 &
0.028
 &
0.017
\\   
18.7
 &
20.03
 &
21.4
 &
0.70
 &
0.79
 &
0.90
 &
0.417
 &
0.010
 &
0.023
 &
0.320
 &
0.014
 &
0.016
\\   
18.7
 &
20.04
 &
21.4
 &
0.90
 &
1.03
 &
1.20
 &
0.406
 &
0.012
 &
0.023
 &
0.280
 &
0.014
 &
0.011
\\   
18.7
 &
20.05
 &
21.4
 &
1.20
 &
1.49
 &
4.00
 &
0.344
 &
0.017
 &
0.021
 &
---
 &
---
 &
---
\\   
21.4
 &
22.88
 &
24.4
 &
0.55
 &
0.62
 &
0.70
 &
0.456
 &
0.015
 &
0.025
 &
0.316
 &
0.015
 &
0.015
\\   
21.4
 &
22.89
 &
24.4
 &
0.70
 &
0.79
 &
0.90
 &
0.446
 &
0.011
 &
0.024
 &
0.312
 &
0.013
 &
0.015
\\
21.4
 &
22.88
 &
24.4
 &
0.90
 &
1.03
 &
1.20
 &
0.382
 &
0.011
 &
0.022
 &
0.277
 &
0.023
 &
0.011
\\   
21.4
 &
22.89
 &
24.4
 &
1.20
 &
1.50
 &
4.00
 &
0.370
 &
0.017
 &
0.022
 &
---
 &
---
 &
---
\\
24.4
 &
26.02
 &
27.7
 &
0.40
 &
0.47
 &
0.55
 &
---
 &
---
 &
---
 &
0.180
 &
0.065
 &
0.009
\\   
24.4
 &
26.02
 &
27.7
 &
0.55
 &
0.62
 &
0.70
 &
0.400
 &
0.016
 &
0.022
 &
0.260
 &
0.016
 &
0.013
\\   
24.4
 &
26.02
 &
27.7
 &
0.70
 &
0.79
 &
0.90
 &
0.407
 &
0.011
 &
0.023
 &
0.283
 &
0.015
 &
0.013
\\   
24.4
 &
26.03
 &
27.7
 &
0.90
 &
1.03
 &
1.20
 &
0.405
 &
0.012
 &
0.023
 &
0.280
 &
0.016
 &
0.011
\\   
24.4
 &
26.04
 &
27.7
 &
1.20
 &
1.51
 &
4.00
 &
0.321
 &
0.014
 &
0.019
 &
---
 &
---
 &
---
\\   
\bottomrule 
\end{tabular} 
\noindent
\begin{tabular}{cccccc|ccc|ccc} 
\toprule 
$p$\ap{min}
 &
$\langle p \rangle$
 &
$p$\ap{max}
 &
$p_{\mathrm{T}}^\text{min}$
 &
$\langle p_{\mathrm{T}} \rangle$
 &
$p_{\mathrm{T}}^\text{min}$
 &
\Rinc
 &
$\delta\ped{uncorr} $
 &
$\delta\ped{corr} $
 &
\Rexc
 &
$\delta\ped{uncorr} $
 &
$\delta\ped{corr} $
\\  \midrule 
27.7
 &
29.52
 &
31.4
 &
0.40
 &
0.47
 &
0.55
 &
---
 &
---
 &
---
 &
0.226
 &
0.059
 &
0.011
\\   
27.7
 &
29.52
 &
31.4
 &
0.55
 &
0.62
 &
0.70
 &
0.349
 &
0.019
 &
0.023
 &
0.224
 &
0.034
 &
0.011
\\   
27.7
 &
29.53
 &
31.4
 &
0.70
 &
0.79
 &
0.90
 &
0.430
 &
0.014
 &
0.024
 &
0.273
 &
0.012
 &
0.014
\\   
27.7
 &
29.53
 &
31.4
 &
0.90
 &
1.03
 &
1.20
 &
0.394
 &
0.012
 &
0.023
 &
0.253
 &
0.011
 &
0.010
\\   
27.7
 &
29.53
 &
31.4
 &
1.20
 &
1.52
 &
4.00
 &
0.357
 &
0.015
 &
0.021
 &
---
 &
---
 &
---
\\   
31.4
 &
35.55
 &
40.0
 &
0.40
 &
0.47
 &
0.55
 &
---
 &
---
 &
---
 &
0.279
 &
0.050
 &
0.017
\\   
31.4
 &
35.56
 &
40.0
 &
0.55
 &
0.62
 &
0.70
 &
0.398
 &
0.017
 &
0.023
 &
0.247
 &
0.032
 &
0.013
\\   
31.4
 &
35.57
 &
40.0
 &
0.70
 &
0.79
 &
0.90
 &
0.412
 &
0.013
 &
0.024
 &
0.257
 &
0.012
 &
0.014
\\   
31.4
 &
35.58
 &
40.0
 &
0.90
 &
1.03
 &
1.20
 &
0.385
 &
0.010
 &
0.022
 &
0.268
 &
0.009
 &
0.009
\\   
31.4
 &
35.60
 &
40.0
 &
1.20
 &
1.52
 &
4.00
 &
0.350
 &
0.010
 &
0.021
 &
---
 &
---
 &
---
\\   
40.0
 &
45.08
 &
50.5
 &
0.55
 &
0.62
 &
0.70
 &
0.326
 &
0.019
 &
0.025
 &
0.261
 &
0.016
 &
0.015
\\   
40.0
 &
45.08
 &
50.5
 &
0.70
 &
0.79
 &
0.90
 &
0.362
 &
0.014
 &
0.022
 &
0.280
 &
0.014
 &
0.016
\\   
40.0
 &
45.08
 &
50.5
 &
0.90
 &
1.03
 &
1.20
 &
0.382
 &
0.013
 &
0.023
 &
0.272
 &
0.037
 &
0.012
\\   
40.0
 &
45.12
 &
50.5
 &
1.20
 &
1.54
 &
4.00
 &
0.364
 &
0.012
 &
0.022
 &
0.229
 &
0.011
 &
0.005
\\   
50.5
 &
76.35
 &
110.0
 &
0.70
 &
0.79
 &
0.90
 &
0.351
 &
0.021
 &
0.025
 &
---
 &
---
 &
---
\\   
50.5
 &
76.50
 &
110.0
 &
0.90
 &
1.03
 &
1.20
 &
0.323
 &
0.013
 &
0.019
 &
---
 &
---
 &
---
\\   
50.5
 &
76.97
 &
110.0
 &
1.20
 &
1.55
 &
4.00
 &
0.336
 &
0.010
 &
0.020
 &
0.244
 &
0.010
 &
0.012
\\  \bottomrule 
\end{tabular} 
\bigskip


\addcontentsline{toc}{section}{References}
\bibliographystyle{LHCb}
\bibliography{main,standard,LHCb-PAPER,LHCb-CONF,LHCb-DP,LHCb-TDR}

\clearpage
\centerline
{\large\bf LHCb collaboration}
\begin
{flushleft}
\small
R.~Aaij$^{32}$\lhcborcid{0000-0003-0533-1952},
A.S.W.~Abdelmotteleb$^{50}$\lhcborcid{0000-0001-7905-0542},
C.~Abellan~Beteta$^{44}$,
F.~Abudin{\'e}n$^{50}$\lhcborcid{0000-0002-6737-3528},
T.~Ackernley$^{54}$\lhcborcid{0000-0002-5951-3498},
B.~Adeva$^{40}$\lhcborcid{0000-0001-9756-3712},
M.~Adinolfi$^{48}$\lhcborcid{0000-0002-1326-1264},
H.~Afsharnia$^{9}$,
C.~Agapopoulou$^{13}$\lhcborcid{0000-0002-2368-0147},
C.A.~Aidala$^{76}$\lhcborcid{0000-0001-9540-4988}
% ...

{\footnotesize \it

% ...
$^{9}$Universit{\'e} Clermont Auvergne, CNRS/IN2P3, LPC, Clermont-Ferrand, France\\
% ...
$^{13}$LPNHE, Sorbonne Universit{\'e}, Paris Diderot Sorbonne Paris Cit{\'e}, CNRS/IN2P3, Paris, France\\
% ...
$^{32}$Nikhef National Institute for Subatomic Physics, Amsterdam, Netherlands\\
% ...
$^{40}$Instituto Galego de F{\'\i}sica de Altas Enerx{\'\i}as (IGFAE), Universidade de Santiago de Compostela, Santiago de Compostela, Spain\\
% ...
$^{44}$Physik-Institut, Universit{\"a}t Z{\"u}rich, Z{\"u}rich, Switzerland\\
% ...
$^{48}$H.H. Wills Physics Laboratory, University of Bristol, Bristol, United Kingdom\\
% ...
$^{50}$Department of Physics, University of Warwick, Coventry, United Kingdom\\
% ...
$^{54}$Oliver Lodge Laboratory, University of Liverpool, Liverpool, United Kingdom\\
% ...
$^{62}$Syracuse University, Syracuse, NY, United States\\
% ...
$^{76}$University of Michigan, Ann Arbor, MI, United States, associated to $^{62}$\\
\bigskip
% ...
}
\end{flushleft}
\end{document}